\documentclass[]{aastex701}
\usepackage{amsmath}
\usepackage{booktabs}
\usepackage{microtype}  
\usepackage{graphicx}
\usepackage{hyperref}
\usepackage{makecell}
\usepackage{amsmath}

\graphicspath{{../data/figures/}}

\begin{document}
\title{$r$-Process as Production of Lanthanides from Compact Object Mergers}

\author[0009-0008-7985-2455]{Li-Ming Zhuo}
\affiliation{Guangxi Key Laboratory for Relativistic 
Astrophysics, School of Physical Science and Technology, Guangxi University, Nanning 530004, People's Republic of China}
\email{2507300159@st.gxu.edu.cn}

\author[0000-0001-6396-9386]{Hou-Jun L\"{u}}
\altaffiliation{Corresponding author: lhj@gxu.edu.cn (LHJ)}
\affiliation{Guangxi Key Laboratory for Relativistic 
Astrophysics, School of Physical Science and Technology, Guangxi University, Nanning 530004, People's Republic of China}
\email{lhj@gxu.edu.cn}

\author[0009-0003-6940-2171]{Xiao-Xuan Liu}
\affiliation{Guangxi Key Laboratory for Relativistic 
Astrophysics, School of Physical Science and Technology, Guangxi University, Nanning 530004, People's Republic of China}
\email{15195809991@163.com}

\author[0000-0001-8406-8683]{Meng-Hua Chen}
\affiliation{Guangxi Key Laboratory for Relativistic 
Astrophysics, School of Physical Science and Technology, Guangxi University, Nanning 530004, People's Republic of China}
\email{mhc@gxu.edu.cn}

\author[0000-0002-7044-733X]{En-Wei Liang}
\affiliation{Guangxi Key Laboratory for Relativistic 
Astrophysics, School of Physical Science and Technology, Guangxi University, Nanning 530004, People's Republic of China}
\email{lew@gxu.edu.cn}

\begin{abstract}
The relative contributions of binary neutron star (BNS) and black hole-neutron star (BH-NS) mergers to the Galactic enrichment from the rapid neutron-capture process ($r$-process) remain an open question in nuclear astrophysics. In this paper, we present end-to-end numerical simulations of binary population synthesis, nucleosynthesis, and Galactic chemical evolution (GCE) via Compact Object Mergers: Population Astrophysics and Statistics (COMPAS), SkyNet network, and a stochastic multizone model, respectively. By considering five equations of state (EOSs) of NS and different spins of BH for BNS and BH-NS mergers, it is found that the averaged abundance seems not to be significantly dependent on the EOS of the neutron star (NS) and the spin of the BH for the BNS and BH-NS mergers. For the given EOSs of NS, the abundance of BNS is higher than that of BH-NS mergers before the second peak (for mass number $A\sim$130), but it is lower when the mass number is greater than 130. Moreover, the lighter $r$-process elements ($A<$130) are predominantly produced by the disk wind, while the heavier elements ($A>$130) originate mainly from the dynamical ejecta. The purely BNS mergers, on the other hand, are the dominant contributors to the production of lanthanides like europium (Eu) at the later stage of GCE, but BH-NS mergers make only a small contribution of heavy $r$-process elements for the $\rm Eu/Fe$ evolutionary tracks in the Milky Way. 
\end{abstract}



\section{Introduction} \label{sec:intro}
The coalescence of two compact stars is expected to produce both gravitational wave (GW) emission and electromagnetic (EM) radiation, such as short duration Gamma-ray bursts (GRBs) and kilonova emission \citep{2014ARA&A..52...43B,2018pgrb.book.....Z}. For example, the first direct detection of GW emission from GW150914 is originated from the merger of a binary black hole \citep[BBH;][]{2016PhRvL.116f1102A}, and an EM counterpart (GRB 170817A and AT2017gfo) associated with a GW event (GW170817) was detected by Laser Interferometer Gravitational Wave Observatory (LIGO) and other space/ground-based telescopes \citep{2017ApJ...848L..12A,2017ApJ...848L..14G,2017ApJ...848L..15S,2018NatCo...9..447Z}. Moreover, several GW events from the merger of black hole and neutron star (BH-NS) have been reported, such as GW200105, GW200115, and GW230529 \citep{2021ApJ...915L...5A,2023PhRvD.108h3018Y,2024ApJ...970L..34A}, and the mergers of neutron star and white dwarf (NS-WD) are also proposed to produce low-frequency GWs and short GRBs \citep{2022Natur.612..232Y,2023ApJ...947L..21Z,2025ApJ...988L..46L}.

More interestingly, the heavier element strontium was identified in the observed spectrum of GRB 170817A/GW170817 \citep{2019Natur.574..497W}, and the later red kilonova emission with a lanthanide-rich component indirectly shows the possibility of the existence of lanthanides \citep{2017ApJ...851L..21V}. It is suggested that binary neutron star (BNS) mergers are major astrophysical sites for the rapid neutron-capture process ($r$-process) to produce heavier elements \citep{2017PhRvL.119p1101A, 2017ApJ...848L..17C, 2017Natur.551...80K, 2017ApJ...848L..18N, 2021ApJ...919...59C}. BH-NS systems are other possible astrophysical sites for the $r$-process to produce heavier elements with a lower contribution efficiency, but it depends on the spin and mass of BH  \citep{2016NatCo...712898J,2021ARNPS..71..433L,2021LRR....24....5K}. Hydrodynamic simulations have shown that the astrophysical conditions in NS-WD mergers are unsuitable for the production of heavy elements  \citep{2019MNRAS.488..259F,2023ApJ...956...71K}, especially for those elements with mass number $A>90$ \citep{2025ApJ...988L..46L}. Furthermore, core-collapse supernovae (CCSNe) are also thought to be a favored production site \citep{1994ApJ...433..229W,1996ApJ...471..331Q}, but they are capable of producing weak $r$-process nuclei \citep{2013JPhG...40a3201A,2013ApJ...770L..22W}.

The lanthanides are composed of the elements from lanthanum (atomic number $Z=57$) to lutetium ($Z=71$). The astrophysical origins of more than half of the elements heavier than iron in the Universe are attributed to the $r$-process \citep{2007PhR...450...97A}. On the other hand, \citet{2014AJ....147..136R} reported the metal abundance for metal-poor stars (MPSs) in the Galactic halo, and found that the distributions and evolutions of $[\mathrm{Fe/H}]$ and $[\mathrm{Eu/Fe}]$ exhibit a pronounced intrinsic dispersion. Based on the higher occurrence rate of a BNS with $10$--$1700\,\mathrm{Gpc}^{-3}\,\mathrm{yr}^{-1}$ measured by Gravitational-Wave Transient Catalog (GWTC)-3 \citep{2023PhRvX..13a1048A}, BNS mergers can serve as the dominant contributor to the production of heavy $r$-process elements in the low metallicity region of the Milky Way, such compact binary mergers have sufficiently short delay times \citep{2023ApJ...943L..12K,2024MNRAS.529.1154C,2025arXiv250806020S}. However, one basic question is whether BNS mergers can self-consistently account for such pronounced abundance dispersion. 

Although progress has been made in previous studies to address the above question, such as in \citet{2020MNRAS.494.4867V}, who traced the inhomogeneous mixing of $r$-process elements by using high-resolution hydrodynamic simulations. \citet{2018ApJ...855...99C} adopted semianalytical simulations with static single-event yields, which inherently erase the intrinsic dispersion of binary parameters present in real populations. \citet{2015MNRAS.452.1970W} successfully reproduced the halo $r$-process abundance dispersion by using a parametrized inhomogeneous Galactic chemical evolution (GCE) model, yet their delay-time and yield prescriptions lacked direct support from binary evolution physics. \citet{2021MNRAS.508.5028B} utilized Compact Object Mergers: Population Astrophysics and Statistics (COMPAS) to explore cosmic merger rates and kinematic properties of BH-NS binaries, but their analysis was largely limited to identifying the fraction of tidally disrupted systems, stopping short of mapping specific dynamic ejecta profiles to GCE nucleosynthetic yields.

In this work, we present a comprehensive framework to couple COMPAS binary population synthesis, SkyNet nucleosynthesis, and a stochastic GCE model (COMPAS $\rightarrow$ SkyNet $\rightarrow$ GCE), and then investigate the $r$-process nucleosynthesis and contributions from BNS and BH-NS mergers by considering five equations of state (EOSs) of NS. This paper is organized as follows. In Section~\ref{sec:compas}, we introduce the detailed characteristics of the initial binary population modeled via COMPAS. The construction of the SkyNet nucleosynthesis grid to produce the heavy elements is presented in Section~\ref{sec:skynet}. The simulation results from the multi-zone stochastic GCE model and analysis of the abundance dispersion trajectories of MPS are shown in Section~\ref{sec:gce}. Conclusions are drawn in Section~\ref{sec:conclusion}, along with some additional discussion.

\section{COMPAS Population Synthesis and its Ejecta Evolution} \label{sec:compas}
COMPAS is a public rapid binary population synthesis code to generate populations of isolated stellar binaries within the parametrized assumptions, and then compare with observational data \citep{2022ApJS..258...34R}. In this Section, we will introduce the details of binary population synthesis and the ejecta evolution by adopting the COMPAS code.

\subsection{COMPAS Population Synthesis}
By obtaining the initial binary star parameters (e.g., primary mass, secondary mass, and orbital semi-major axes) from COMPAS \citep{2022ApJS..258...34R} and setting the environmental metallicity to the solar abundance of $0.0142$, we initialize $10^6$ zero-age main sequence (ZAMS) binary star systems by adopting the Kroupa initial mass function (IMF) \citep{2001MNRAS.322..231K}. During binary evolution, the common envelope (CE) phase is driven by the standard $\lambda \alpha_{\mathrm{CE}}$. Here, $\alpha_{\mathrm{CE}}$ is an efficiency factor to represent the fraction of the released orbital energy used to unbind the envelope, and $\lambda$ is a structure parameter characterizing the envelope's binding energy \citep{2022ApJS..258...34R}. The gravitational binding energy ($E_{\mathrm{bind}}$) of the envelope is expressed as
\begin{equation}
E_{\mathrm{bind}} = G m_1 M_{\mathrm{env}} / (\lambda R_1),
\label{eq:gbe}
\end{equation}
and the energy loss from orbital contraction ($\Delta E_{\mathrm{orb}}$) is given by 
\begin{equation}
\Delta E_{\mathrm{orb}} = G M_c m_2/(2a_f) - G m_1 m_2/(2a_i) > 0.
\label{eq:orb}
\end{equation}
Here, $m_1$, $m_2$, $M_c$, and $M_{\mathrm{env}}=m_1-M_{\rm c}$ are the primary star mass, secondary star mass, core mass of the primary star, and the envelope mass of the primary star, respectively. $a_i$ and $a_f$ are the initial and final orbital semi-major axes of the CE phase, respectively. $R_1$ is the radius of the primary star, and $G$ is the gravitational constant.

Based on energy conservation, the gravitational binding energy of the envelope should be equal to energy loss from orbital contraction by considering the efficiency factor $\alpha_{\mathrm{CE}}$ \citep{1984ApJ...277..355W}, namely, 
\begin{equation}
E_{\mathrm{bind}} = \frac{G m_1 M_{\mathrm{env}}}{\lambda R_1}
= \alpha_{\mathrm{CE}} \left( \frac{G M_c m_2}{2 a_f} - \frac{G m_1 m_2}{2 a_i} \right).
\label{eq:ce}
\end{equation}
After the CE phase, the energy loss of the binary system is dominated by GW radiation, and it can result in orbital radius getting closer and closer \citep{1964PhRv..136.1224P}. The GW delay-time distribution (DTD) follows a power-law distribution, namely, $\frac{dN}{dt} \propto t^{\sim -1.1}$ for $t>1$ Gyr, where $N$ is the number of merger events \citep{2019MNRAS.487.4847B}. One needs to note that the number of binary star systems should be increased if we adopt a smaller value of environmental metallicity, namely, $Z<0.0142$, but $dN/dt$ is still obeyed $t^{-1.1}$, namely, $\frac{dN}{dt} \propto t^{\sim -1.1}$ which does not effect our results after normalization of time \citep{2015MNRAS.451.4086S,2019MNRAS.490.3740N}.

\begin{figure*}[ht!]
\centering
\includegraphics[width=0.48\linewidth]{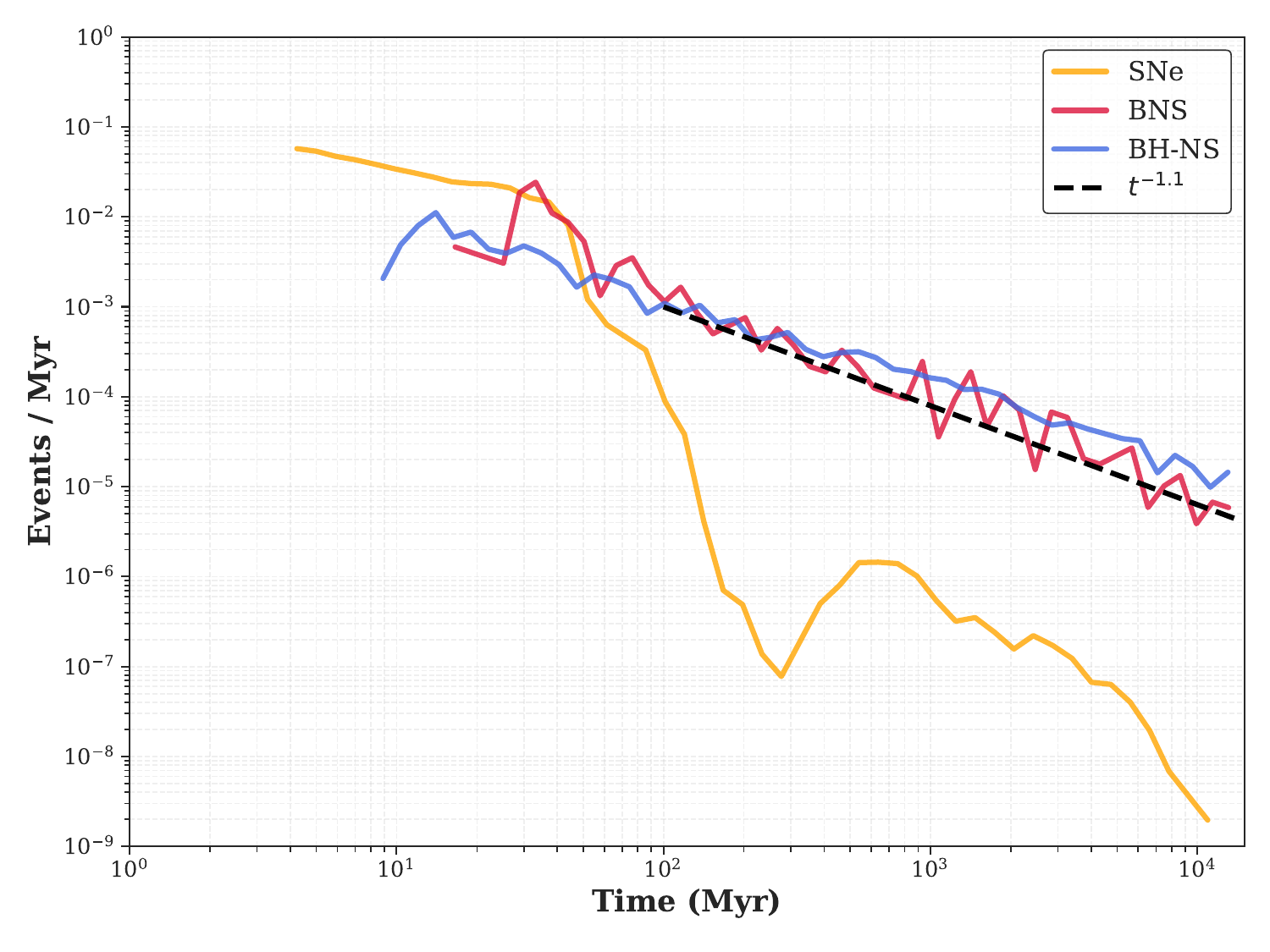}
\includegraphics[width=0.45\linewidth]{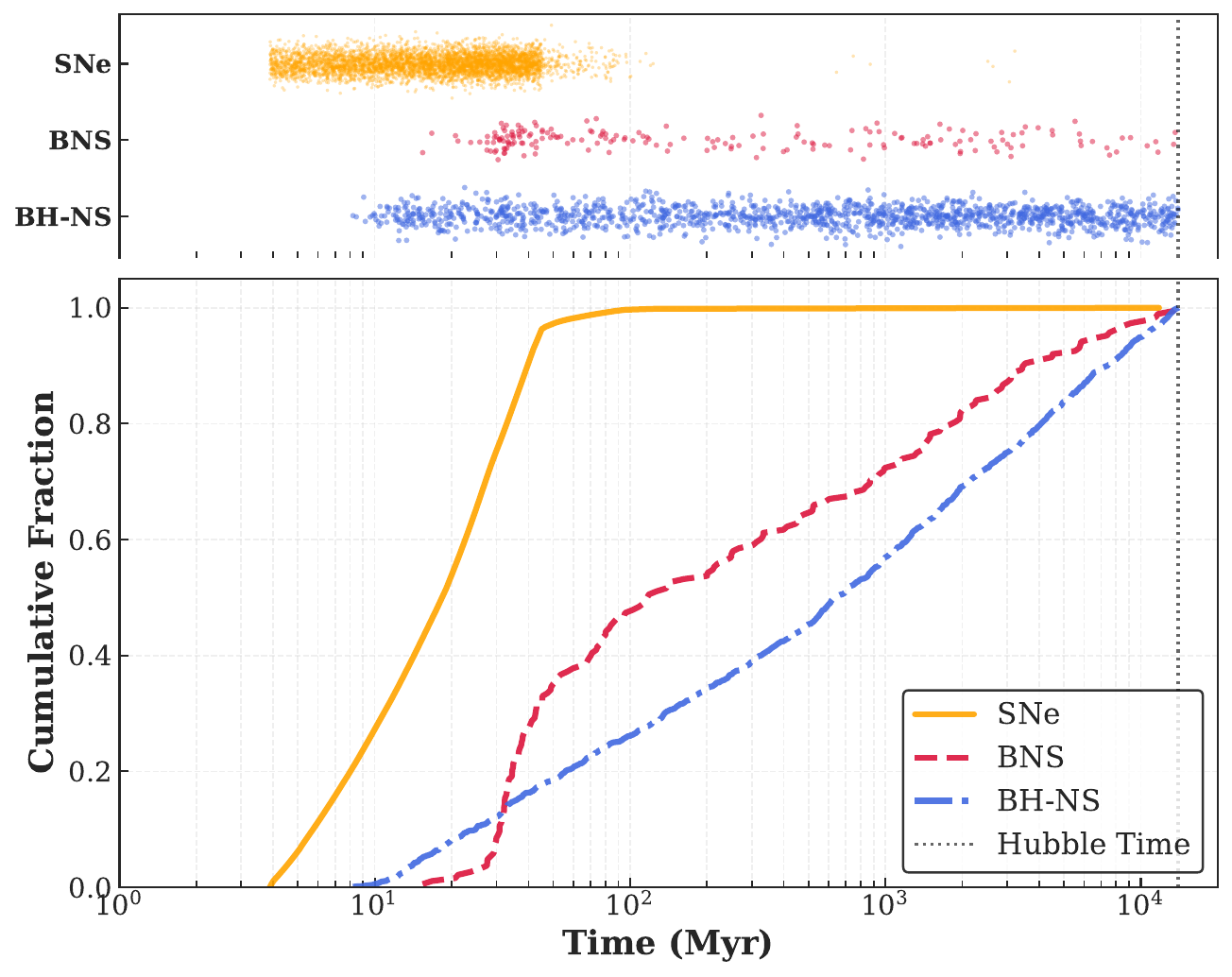}
\caption{Normalized DTD for the SNe, BNS, and BH-NS channels generated by COMPAS. Left panel: the event rate of coalescence as function of time for BNS (red solid curve) and BH-NS (blue solid curve), along with the explosion rate of SNe (yellow solid curve). The black dashed line is the power-law function with canonical $t^{-1.1}$ driven by gravitational-wave radiation. Right panel (top): distributions of simulated individual events for SNe (orange points), BNS (pink points), and BH-NS (blue points) as function of time. Right panel (bottom): Cumulative fraction as a function of time for SNe (yellow solid curve), BNS (red dashed curve), and BH-NS (dashed bule curve), and the vertical dotted line is the Hubble time ($t_{\rm H} \sim 14$ Gyr)}.
\label{fig:1}
\end{figure*}

\begin{figure*}[ht!]
\centering
\begin{minipage}{0.48\linewidth}
    \centering
    \includegraphics[width=\linewidth]{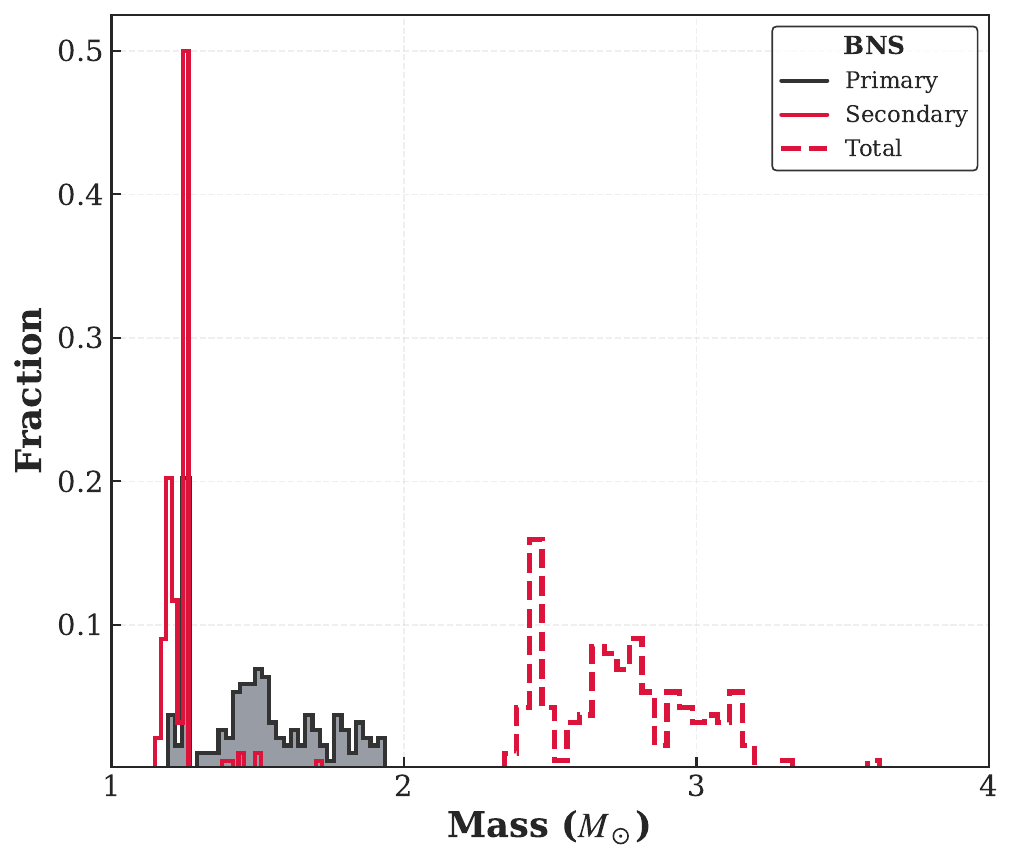}
\end{minipage}
\hfill
\begin{minipage}{0.48\linewidth}
    \centering
    \includegraphics[width=\linewidth]{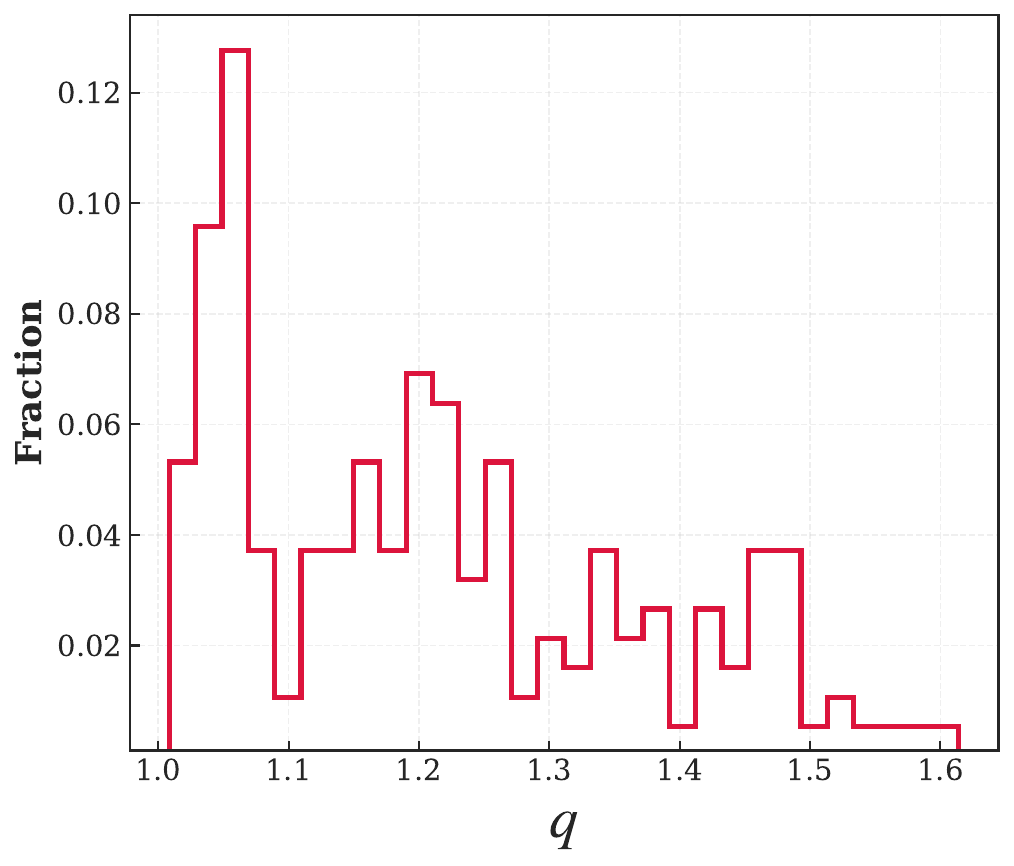}
\end{minipage}

\vspace{0.3cm}

\begin{minipage}{0.48\linewidth}
    \centering
    \includegraphics[width=\linewidth]{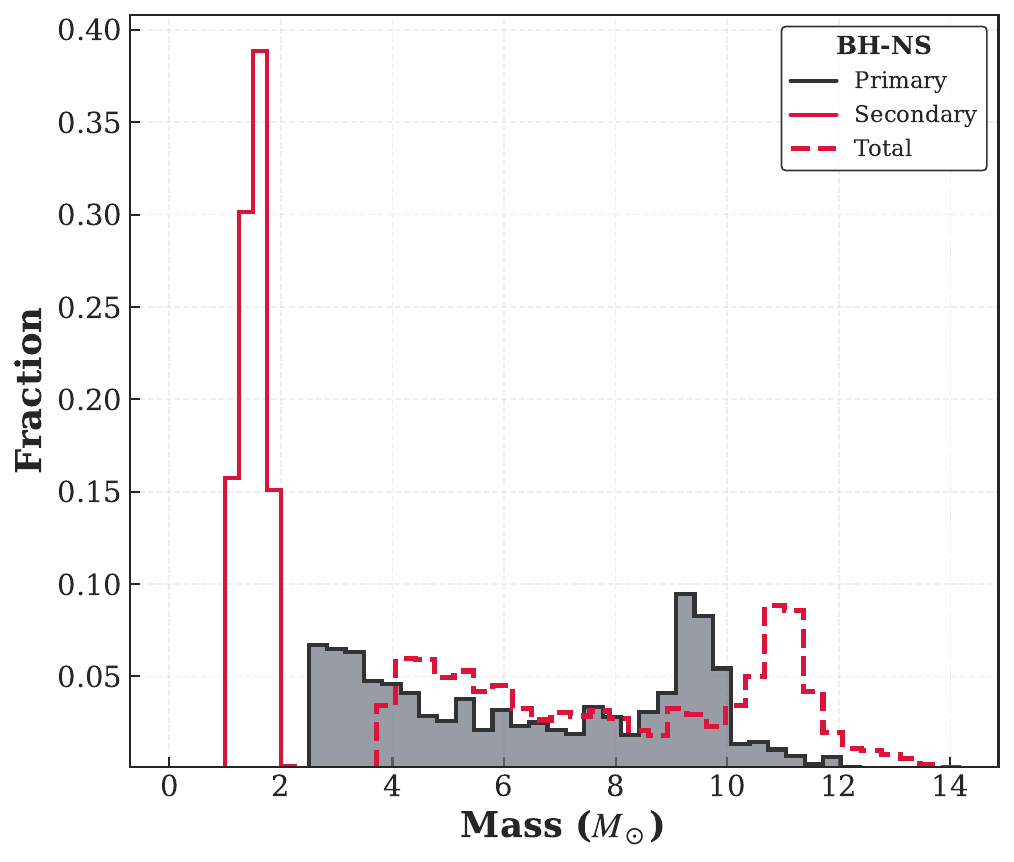}
\end{minipage}
\hfill
\begin{minipage}{0.48\linewidth}
    \centering
    \includegraphics[width=\linewidth]{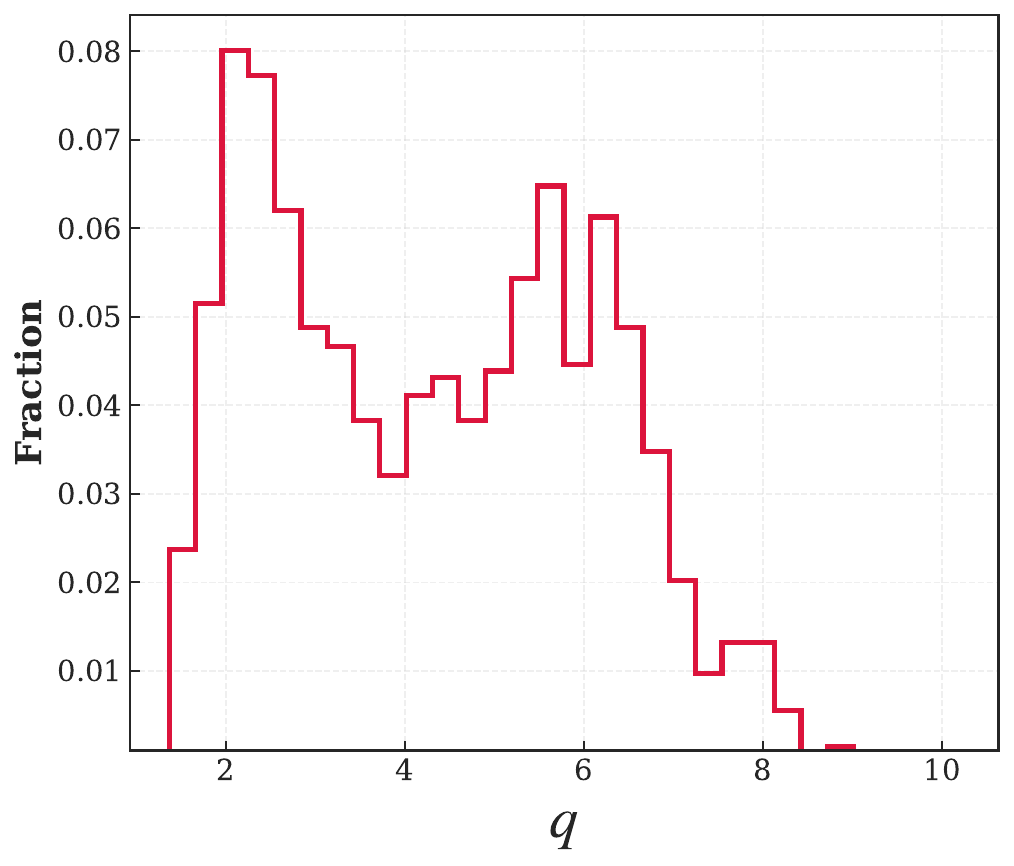}
\end{minipage}

\caption{Distributions of primary mass, secondary mass, total mass, and mass ratio for BNS and BH-NS populations synthesized via COMPAS.}
\label{fig:2}
\end{figure*}

Then, the evolution timescale is extended up to the Hubble time ($t_{\rm H}$), and the DTD is integrated from $t=$0.035 Gyr to the Hubble time \citep{2019MNRAS.487.4847B}. Physically, the occurrence of a supernova explosion the formation of a compact star first require the prior formation and evolution of a star. Thus, $t=$0.035 Gyr roughly corresponds to the average evolutionary lifetime from ZAMS to CCSNe, and then further to form compact objects. Since the timescale of evolution for ZAMS is roughly in a range from $t=$0.004 Gyr to $t=$0.05 Gyr \citep{2019MNRAS.487.4847B}, so that we adopt $t=$0.035 Gyr as the average of evolutionary lifetime. The normalization condition decouples the DTD shape from the merger efficiency ($\eta$) in Section~\ref{sec:gce}, namely, the DTD shape determines the temporal distribution of enrichment events, while $\eta$ dictates the scale of the total number of events. We extract the SNe and double the compact object systems (including BNS and BH-NS) that successfully merge within this time window. One can obtain the total number of SNe ($N=$558880), BNS ($N=$188), and BH-NS ($N=$1436). The normalized DTDs for SNe, BNS, and BH-NS generated by COMPAS are shown in Figure~\ref{fig:1}. It is found that most SNe ($\sim 97.3\%$) are concentrated at $<50\,\mathrm{Myr}$, and about $47.3\%$ (or $26.0\%$) of BNS (or BH-NS) are distributed $<100\,\mathrm{Myr}$. Moreover, Figure~\ref{fig:2} shows the mass distributions of primary stars and secondary stars in BNS and BH-NS systems, respectively. We also present the total mass ($M=m_1+m_2$) and mass ratio ($q=m_1/m_2$) distributions of BNS and BH-NS systems. It is found that the mass ratio of BNS systems predominantly exhibits near-equal mass ratios ($q \sim 1$) with masses tightly clustered around typical neutron star values. In contrast, the mass ratio of BH-NS systems displays a broader range up to $q \sim 10$.

\subsection{Equation of State of NS and Spin of Black Hole}
Based on the results from COMPAS population synthesis, one can obtain the mass distributions of NS in BNS and BH-NS systems. However, the compactness of NS ($C_{\rm NS} = GM_{\rm NS}/Rc^2$) and ejecta properties highly depend on the maximum mass and nuclear EOS which are currently not well constrained  \citep{2014PhRvD..89d7302L,2015ApJ...805...89L,2016PhRvD..94h3010L,2016PhRvD..93d4065G,2020ApJ...890...99L,2025ApJ...980..220H}. Due to the poorly known EOS of NS, we adopt five representative EOSs from the nuclear physics literature \citep{1998PhRvC..58.1804A, 1996ApJ...469..794E, 2010PhRvC..81a5803T, 2013ApJ...774...17S, 2014ApJS..214...22B}. The key physical characteristics of these EOSs, including the maximum NS mass $M_{\mathrm{max}}$, typical radius $R$, compactness $C_{\rm NS}$, and tidal deformability parameter $\Lambda$, are summarized in Table~\ref{tab:eos_params}.

On the other hand, even more than 100 events of BBH mergers were reported by LIGO, Virgo, and KAGRA (LVK Collaboration) during the first three observing runs \citep{2023PhRvX..13a1048A}, and the spin of BH ($\chi_{\mathrm{BH}}$) seems to follow a bimodal distribution with high spin and low spin, which is used to distinguish whether BHs are formed through hierarchical mergers or stellar explosions \citep{2017ApJ...840L..24F,2021NatAs...5..749G,2024PhRvL.133e1401L}. However, due to limited observations of BH-NS merger events and the possible tidal spin-up effect of BH \citep{2021A&A...647A.153B}, it is difficult to understand the distribution of the spin of BH in the BH-NS merger systems. Therefore, we adopt three spin ranges (e.g., high spin, medium spin, and low spin) as the initial spin of BH for BH-NS merger systems. The high spin, medium spin, and low spin of BH are assumed to obey a normal distribution with mean values and standard deviation as $\mu=0.8$, 0.4, 0.0 and $\sigma=0.2$, respectively. It should be noted that the native spin evolution model in COMPAS is based on tidal synchronization mechanisms in isolated binaries, which typically predicts $\chi_{\mathrm{BH}} \lesssim 0.2$. However, dynamical environments such as dense star clusters or active galactic nucleus (AGN) disks can produce second-generation BHs through hierarchical mergers, whose spins can reach $\chi_{\rm BH} \sim 0.7$ \citep{2021NatAs...5..749G, 2021MNRAS.505..339M}. The dynamical channels are important astrophysical scenarios currently not covered by isolated binary evolution codes. Our artificial injection strategy is designed phenomenologically to include the contribution of such highspin BHs to the Galactic $r$-process budget.

\begin{table*}[ht!]
\centering
\caption{Maximum mass, radius, compactness, and tidal deformability of NS for five different EOSs. \label{tab:eos_params}}
\begin{tabular*}{\textwidth}{@{\extracolsep{\fill}}lcccccc@{}}
\toprule
EOS Model & $M_{\mathrm{max}}\,[M_\odot]$ & $R_{\mathrm{1.4}}\,[km]$ & $C_{\mathrm{1.4}}$ & $\Lambda_{\mathrm{1.4}}$  \\
\midrule
DD2   & 2.42 & 13.2 & 0.156 & 680 \\
BHBLp & 2.11 & 13.2 & 0.156 & 730 \\
ENG   & 2.24 & 12.1 & 0.170 & 390 \\
SFHo  & 2.06 & 11.9 & 0.173 & 330 \\
APR4  & 2.16 & 11.4 & 0.181 & 260 \\
\bottomrule
\end{tabular*}
\end{table*}

\subsection{Ejecta evolution and analytical formula}
The properties of the ejecta are determined by analytical formulae which are calibrated against numerical relativity simulations. In our calculations, the gravitational mass ($m_{\mathrm{gra}}$) of NS is first converted to baryonic mass $M_{\mathrm{bar}}$ \citep{2001ApJ...550..426L}, namely,
\begin{equation}
M_{\mathrm{bar}} = m_{\mathrm{gra}} + 0.075\,m_{\mathrm{gra}}^2,
\label{eq:bar_mass}
\end{equation}
where $m_{\mathrm{gra}}$ and $M_{\mathrm{bar}}$ are in units of $M_\odot$.

(1). Ejecta mass and velocity of BNS: The dynamical ejecta mass $M_{\mathrm{dyn}}^{\mathrm{BNS}}$ of BNS adopts the multidimensional fitting formula from \citet{2022CQGra..39a5008N} and \citet{2025ApJ...985...42D},
\begin{equation}
\begin{split}
\frac{M_{\mathrm{dyn}}^{\mathrm{BNS}}}{10^{\mathrm{-3}}M_\odot}&= \bigg[a_1\left(\frac{m_2}{m_1}\right)^{1/3}\frac{1-2C_{\mathrm{NS_1}}}{C_{\mathrm{NS_1}}} + b_1\left(\frac{m_2}{m_1}\right)^{\mathrm{l_1}}+ c_1\left(1-\frac{m_1}{M_{\mathrm{bar},NS_1}}\right)\bigg] M_{\mathrm{bar},NS_1} \\
&\quad + (1\leftrightarrow 2) + d_1,
\end{split}
\label{eq:nedora_dyn}
\end{equation}
where the coefficients $a_1=-0.1004$, $b_1=-0.4403$, $c_1=-0.6452$, $d_1=0.2696$, and $l_1=0.3222$ are fixed by \citet{2022CQGra..39a5008N}. The velocity of dynamical ejecta is given by
\begin{equation}
\begin{split}
v_{\mathrm{dyn}}^{\mathrm{BNS}}/c &= \bigg[a_v\left(\frac{m_1}{m_2}\right)\left(1 + c_v C_{\mathrm{NS_1}}\right)\bigg] + (1\leftrightarrow 2) + b_v
\end{split}
\label{eq:nedora_v}
\end{equation}
with constant coefficients $a_v=-0.5631$, $b_v=1.109$, and $c_v=-1.186$ \citep{2022CQGra..39a5008N}. For the mass of remnant disk $M_{\mathrm{disk}}^{\mathrm{BNS}}$, We employ the log-linear fitting formula which is proposed by \citet{2020PhRvD.101j3002K} and calibrated by \citet{2022CQGra..39a5008N},
\begin{equation}
\begin{split}
\log_{10}\left(\frac{M_{\mathrm{disk}}^{\mathrm{BNS}}}{M_\odot}\right) &= \log_{10}(m_1) + \max\bigg[-3.3, c_2\log_{10}(a_2 C_{{\mathrm{NS_1}}} + b_2)\bigg],
\end{split}
\label{eq:kf20_disk}
\end{equation}
where $a_2=-7.184$, $b_2=1.303$, and $c_2=1.613$. For BNS mergers, whether the central remnant is an NS or a prompt black hole is determined based on the EOS-dependent mass threshold $M_{\mathrm{thr}} = k_{\mathrm{thr}} \cdot M_{\mathrm{max}}$ from \citet{2013ApJ...773...78B}, where $k_{\mathrm{thr}} = 2.38 - 3.606\,C_{\mathrm{1.6}}$ and $M_{\mathrm{max}}$ is maximum mass of NS, The mass fraction of disk wind is drawn from a truncated normal distribution ($\mu=0.30$, $\sigma=0.08$, truncated at $[0.15, 0.45]$) to reflect the physical range of magnetically and viscously driven outflows \citep{2016ARNPS..66...23F}.

\begin{figure}[ht!]
\centering
\includegraphics[width=1.0\linewidth]{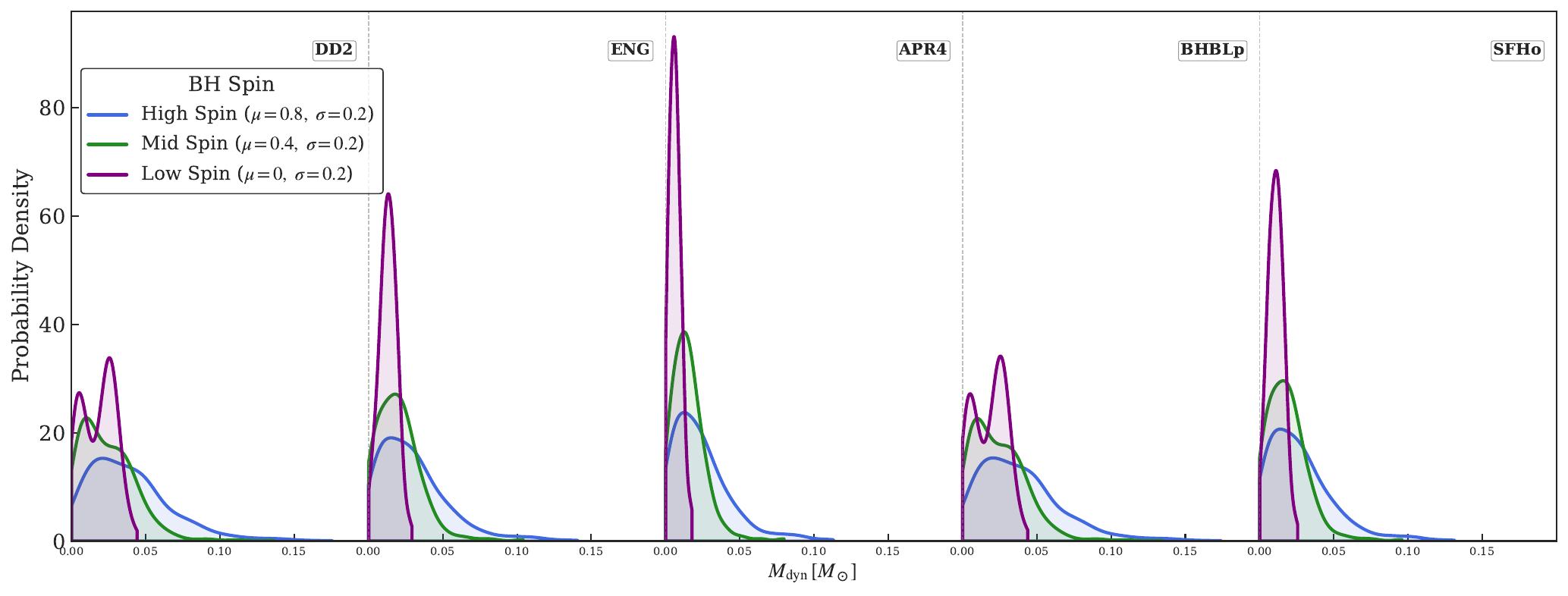}
\caption{Probability density of ejecta mass of BH-NS with different spin of BH for five different EOSs of NS (e.g., DD2, ENG, APR4, BHBLp, and SFHo). In each panel, the curves with shaded regions are the probability distributions of dynamical ejecta mass with low spin (purple), middle spin (green), and high spin (blue) of BH}.
\label{fig:3}
\end{figure}

(2). Ejecta mass and velocity of BH-NS: The total baryonic mass that left outside the event horizon is determined by \citet{2018PhRvD..98h1501F},
\begin{equation}
\begin{split}
\frac{M_{\mathrm{rem}}^{\mathrm{BH-NS}}}{M_{\mathrm{bar},\mathrm{NS}}} &= \max\bigg[\bigg(\alpha_1\frac{1-2C_{\mathrm{NS}}}{P^{1/3}} - \beta_1\frac{R_{\mathrm{ISCO}} C_{\mathrm{NS}}}{m_{\mathrm{BH}} P} + \gamma_1\bigg)^{\delta},\ 0\bigg]
\end{split}
\label{eq:foucart_rem}
\end{equation}
where $\alpha_1=0.406$, $\beta_1=0.139$, $\gamma_1=0.255$, and $\delta=1.761$ are fixed, and $P = q/(1+q)^2$ is the symmetric mass ratio. Here, the dimensionless innermost stable circular orbit (ISCO) radius in the Kerr metric \citep{1972ApJ...178..347B}, $R_{\mathrm{ISCO}}/m_{\mathrm{BH}}$, is given by
\begin{align}
\frac{R_{\mathrm{ISCO}}}{m_{\mathrm{BH}}} &= 3 + Z_2 - \mathrm{sign}(\chi_{\mathrm{BH}})\sqrt{(3 - Z_1)(3 + Z_1 + 2Z_2)} \label{eq:isco} \\[1ex]
Z_1 &= 1 + (1 - \chi_{\mathrm{BH}}^2)^{1/3} \times \big[ (1 + \chi_{\mathrm{BH}})^{1/3} + (1 - \chi_{\mathrm{BH}})^{1/3} \big] \nonumber \\
Z_2 &= \sqrt{3\chi_{\mathrm{BH}}^2 + Z_1^2}. \label{eq:isco_z}
\end{align}
So that, the dynamical ejecta mass is given by \citet{2020PhRvD.101j3002K},
\begin{equation}
\begin{split}
\frac{M_{\mathrm{dyn}}^{\mathrm{BH-NS}}}{M_{\mathrm{bar},\mathrm{NS}}} &=\, \max\bigg[ \alpha_2 q^{n_1}\frac{1-2C_{\mathrm{NS}}}{C_{\mathrm{NS}}} - \beta_2 q^{n_2} \frac{R_{\mathrm{ISCO}}}{m_{\mathrm{BH}}} + \gamma_2,\ 0\bigg],
\end{split}
\label{eq:kf20_dyn}
\end{equation}
where the fitting coefficients are $\alpha_2=0.007116$, $\beta_2=0.001436$, $\gamma_2=-0.02762$, $n_1=0.8636$, and $n_2=1.6840$. Negative raw values represent no dynamical mass ejected ($M_{\mathrm{dyn}}^{\mathrm{BH-NS}}=0$), implying the neutron star plunges directly into the black hole horizon without tidal disruption. Within the different spin scenarios, the probability density distributions of the dynamical ejecta masses are detailed in Figure~\ref{fig:3}. We find that the higher spin of the BH in the BH-NS merger systems corresponds to larger dynamical ejecta mass. Physically, the dynamical ejecta mass of BH-NS merger is dependent on both the radius of tidal disruption and the innermost stable circular orbit (ISCO). For a low-spin BH in the BH-NS system, the tidal disruption radius lies inside the ISCO and the event horizon, and the NS crosses the event horizon and plunges whole into the singularity before appreciable tidal disruption can occur. So that, both the remnant disk mass and the amount of dynamical ejecta are very small or negligible. However, for high-spin BH in the BH-NS system, the frame dragging can contract the event horizon and shift the ISCO inward for prograde orbits. If this is the case, the NS is disrupted far outside the event horizon forming the dynamical ejecta. The velocity of dynamical ejecta is estimated by adopting the formula from \citet{2017CQGra..34d4002F},
\begin{equation}
v_{\mathrm{dyn}}^{\mathrm{BH-NS}}/c = 0.0149 \, q + 0.1493.
\label{eq:fc17_vel}
\end{equation}
Moreover, the mass of the accretion disk of BH-NS is then estimated as $M_{\mathrm{disk}}^{\mathrm{BH-NS}} = M_{\mathrm{rem}}^{\mathrm{BH-NS}} - M_{\mathrm{dyn}}^{\mathrm{BH-NS}}$. The mass loss caused by the BH-NS disk wind can be estimated as \citet{2020MNRAS.497.3221F},
\begin{equation}
M_{\mathrm{wind}}^{\mathrm{BH-NS}} = 0.18 + \frac{0.11}{1 + \exp\left[1.5\,(q - 3)\right]}M_{\mathrm{disk}}^{\mathrm{BH-NS}}
\label{eq:BH-NS_wind}
\end{equation}

(3). Electron fraction ($Y_e$): For BNS mergers, the dynamical ejecta exhibit a multicomponent $Y_e$ structure which reflects the distinct thermodynamic histories of the tidal and shock-heated components \citep{2018ApJ...869..130R}. Therefore, we adopt the results from numerical simulations with a normal distribution \citep{2024MNRAS.529.1154C} as the electron fraction of dynamical ejecta $Y^{\rm BNS}_{e,\mathrm{dyn}}$ =0.17($\sigma=0.04$, truncated at $[0.10, 0.24]$). 
On the other hand, for the electron fraction of the wind ($Y^{\rm BNS}_{e,\mathrm{wind}}$), it depends on the remnant of the BNS merger, which is an NS or a BH. We adopt the results from numerical simulations with a normal distribution, divided into equatorial and polar regions. The electron fractions are set to $Y^{\rm BNS}_{e,\mathrm{wind-polar}}=0.27$ ($\sigma=0.02$, truncated at $[0.24, 0.32]$) and $Y^{\rm BNS}_{e,\mathrm{wind-eq}}= Y^{\rm BNS}_{e,\mathrm{dyn}} $, with a polar mass fraction of $f_{\text{polar}} = 0.30$ for an NS remnant \citep{2017ApJ...851L..21V,2015MNRAS.448..541J}, and $Y^{\rm BNS}_{e,\mathrm{wind}}=0.20$ ($\sigma=0.015$, truncated at $[0.18, 0.22]$) for a BH remnant \citep{2015PhRvD..91l4021F,2023PhRvD.107l3001H}.

For BH-NS mergers, by systematically lacking a collision interface and neutrino source, the dynamical ejecta is produced purely by cold tidal disturbance rather than shock heating, and it retains the pristine, extreme neutron-richness of the original NS \citep{2015PhRvD..91l4021F,2023PhRvD.107l3001H}. We adopt $Y_{e,\mathrm{dyn}}^{\mathrm{BH-NS}} = 0.05$ with a normal distribution ($\sigma=0.01$, truncated at $[0.03, 0.07]$) in our calculations. Moreover, the remnant of the BH-NS merger must be a BH. In the absence of external neutrino irradiation, internal viscous heating and magnetic turbulence become the primary mechanisms driving weak interactions \citep{2015MNRAS.449..390F,2023PhRvD.107l3001H}. The electron fraction of wind remains fundamentally bounded by the lack of neutrino irradiation, and we still adopt $Y_{e,\mathrm{wind}}^{\mathrm{BH-NS}} = 0.20$ ($\sigma=0.03$ truncated at $[0.18, 0.25]$) in our calculations. One needs to note that higher spin of BH shrinks the ISCO and substantially increases the disk mass, which in turn drives stronger viscous heating. However, this slight compositional shift is negligible by compared to the magnitude of the mass enhancement itself. Therefore, the dominant effect of high spin on the $r$-process lies entirely in amplifying the total quantity of ejecta rather than altering its composition.

\section{SkyNet Nucleosynthesis} \label{sec:skynet}

SkyNet is a public nuclear reaction network code which has been widely used to study the evolution of nucleosynthesis in diverse astrophysical scenarios, including $r$-process nucleosynthesis \citep{2017ApJS..233...18L}. The network contains 7843 nuclide species and 140,000 nuclear reactions, and the nuclear reaction rates, nuclear masses, and radioactive decay data are taken from the JINA REACLIB database (\citealt{2010ApJS..189..240C}), AME2020  (\citealt{2021ChPhC..45c0003W}), and NUBASE2020 (\citealt{2021ChPhC..45c0001K}), respectively. \citep{2025ApJ...988L..46L} investigated the heavy elements produced in NS–WD merger nucleosynthesis via SkyNet, and found that the mass number of the heaviest element produced for such an NS-WD merger is less than 90 ($A<90$). In this Section, we also adopt the numerical simulations of nucleosynthesis via SkyNet nuclear reaction network (\citealt{2017ApJS..233...18L}) to produce the heavier elements for both BNS and BH-NS mergers, and then compare with the abundance of the solar system, especially in the range of heavier elements. 

The first step is to construct a grid including 80 physical nodes covering $Y_e \in [0.01, 0.40]$, and each $Y_e$ is computed at two entropies (e.g., $s \in \{10, 20\}$) by using a fixed typical expansion timescale $\tau=7.1\,\mathrm{ms}$. The dynamical ejecta of shock-heated and disk wind of viscous heating are mapped to $s=20$ and $s=10$, respectively (\citealt{2018ApJ...869..130R}). The total ejecta mass can be expressed as
\begin{equation}
\begin{split}
M^{\rm ej}_{\mathrm{total}} &= M^{\rm ej}_{\mathrm{dyn}} \cdot X_i(Y^{\rm ej}_{e,\mathrm{dyn}}, s=20) + M^{\rm ej}_{\mathrm{wind}} \cdot X_i(Y^{\rm ej}_{e,\mathrm{wind}}, s=10)
\end{split}
\label{eq:abs_yield}
\end{equation}
where $X_i = Y_i \cdot A_i$, and $A_i$ and $Y_i$ are the atomic mass number of the $i$th nuclide and the corresponding abundance, respectively. Then, we calculate the average abundance for binary systems (e.g., BNS and BH-NS) based on its dynamical and wind masses with the corresponding abundance, 
\begin{equation}
Y_{\mathrm{avg}}(A) = \frac{1}{N} \sum_{j=1}^{N} 
\frac{M^{\rm ej}_{\mathrm{dyn},j} \cdot Y^{\rm ej}_{\mathrm{dyn},j}(A) + M^{\rm ej}_{\mathrm{wind},j} \cdot Y^{\rm ej}_{\mathrm{wind},j}(A)}
{M^{\rm ej}_{\mathrm{dyn},j} + M^{\rm ej}_{\mathrm{wind},j}}.
\label{eq:unweighted}
\end{equation}
More details of the SkyNet nucleosynthesis can be found in our previous works of \citep{2024MNRAS.529.1154C} and \citep{2025ApJ...988L..46L}.

\begin{figure}[ht!]
\centering
\includegraphics[width=0.7\linewidth]{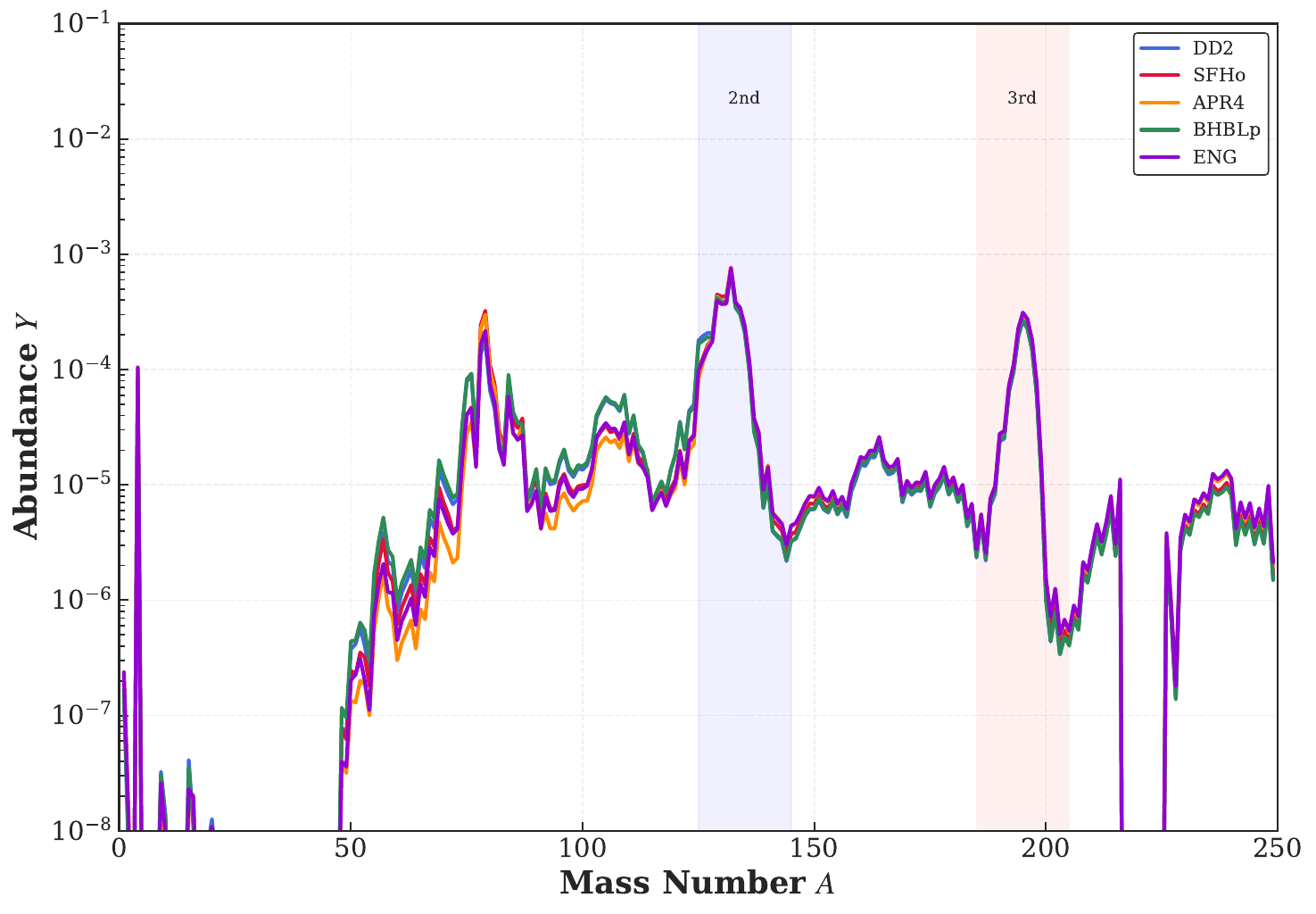}
\caption{Event-averaged abundances of $r$-process for five different EOSs of NS from BNS mergers.}
\label{fig:4}
\end{figure}

\begin{figure}[ht!]
\includegraphics[width=0.48\linewidth]{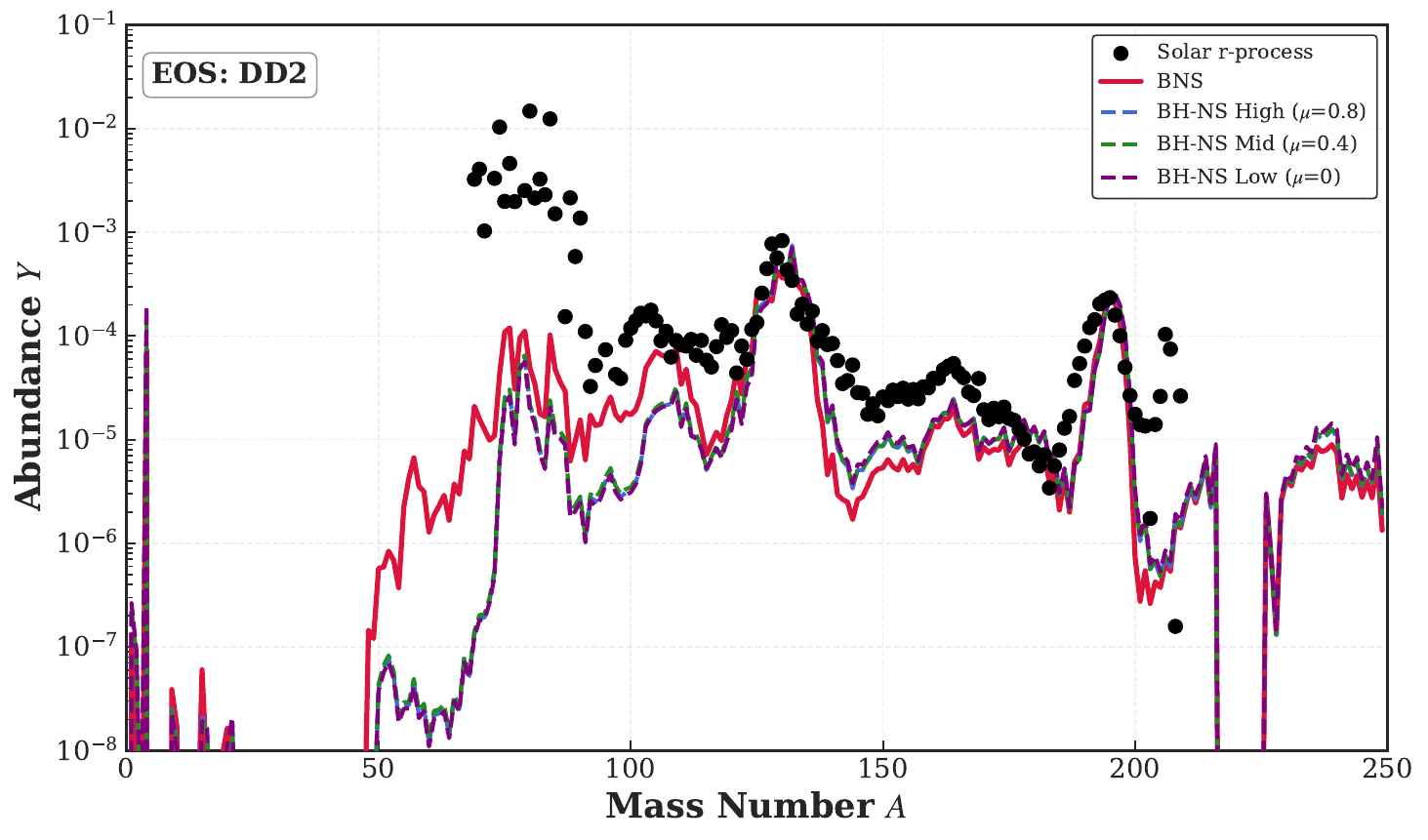}\hfill
\includegraphics[width=0.48\linewidth]{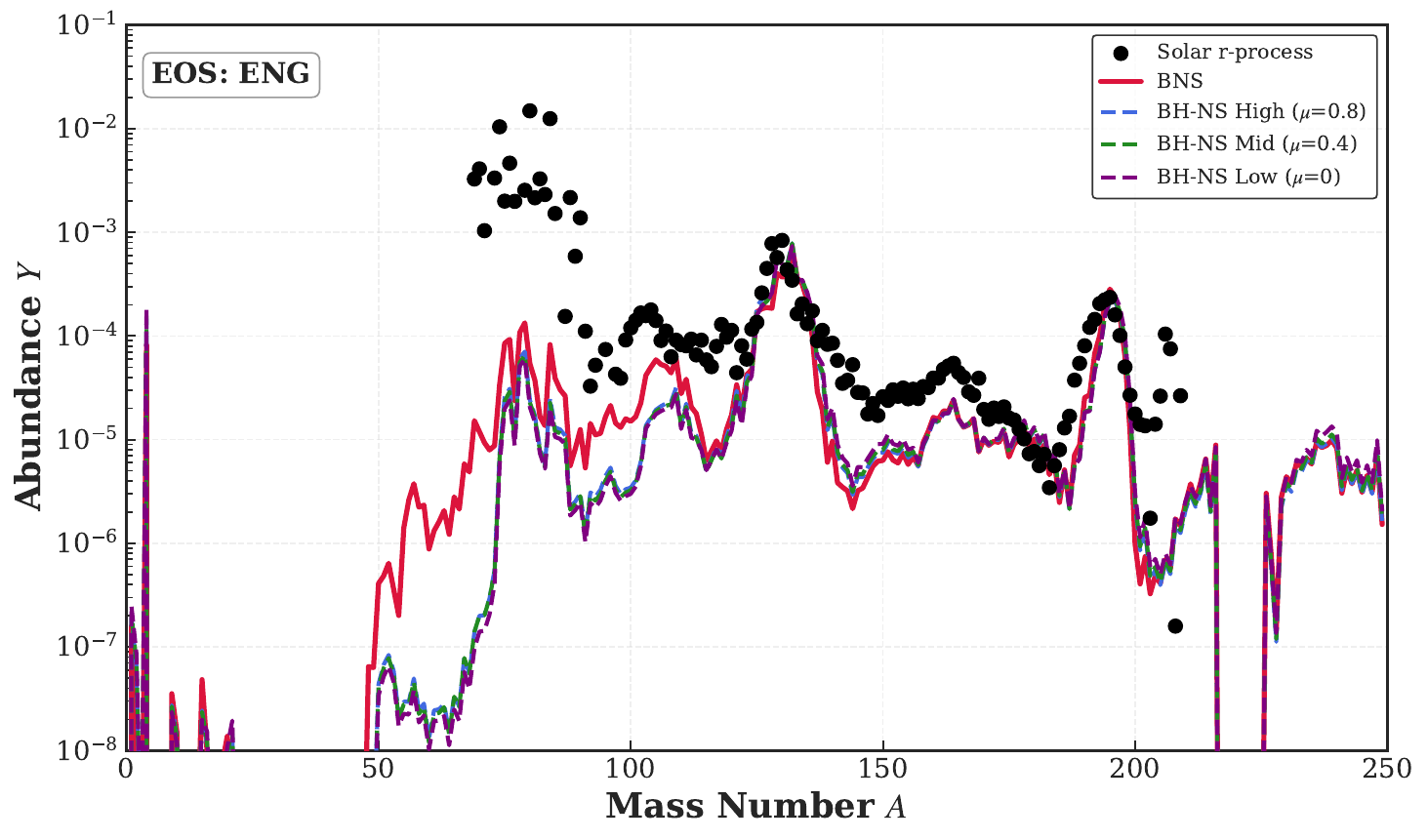}\hfill
\includegraphics[width=0.48\linewidth]{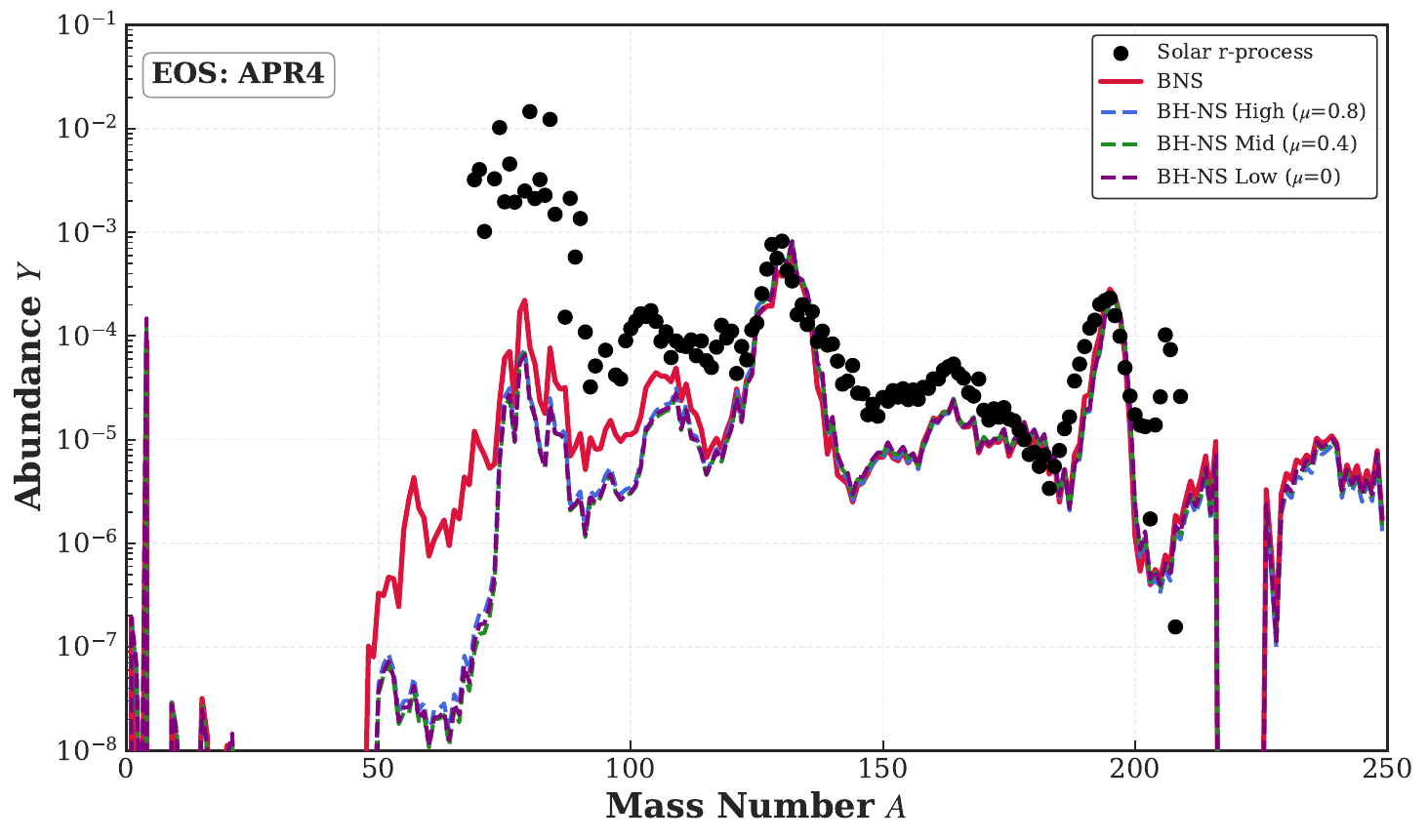}
\includegraphics[width=0.48\linewidth]{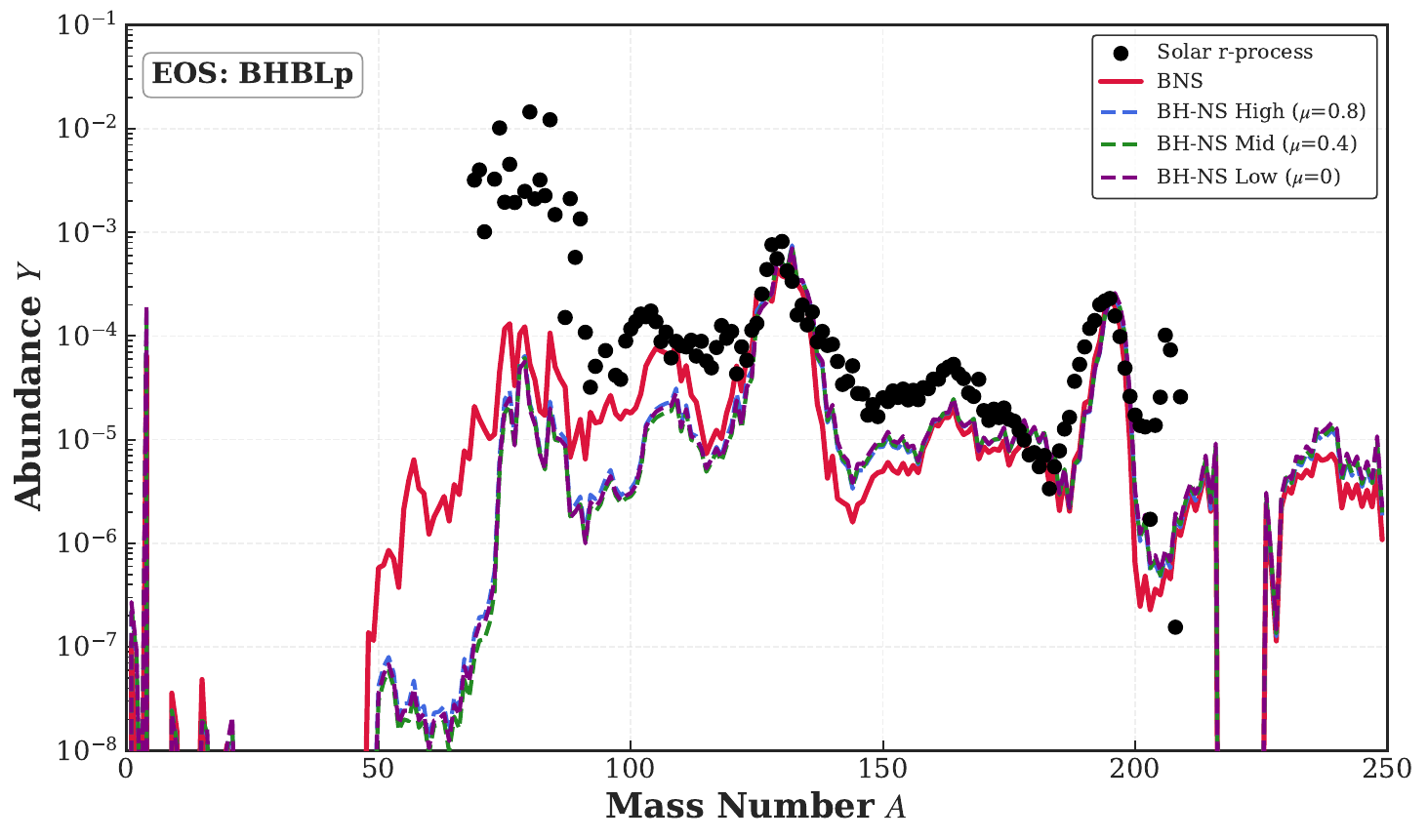}\hfill
\includegraphics[width=0.48\linewidth]{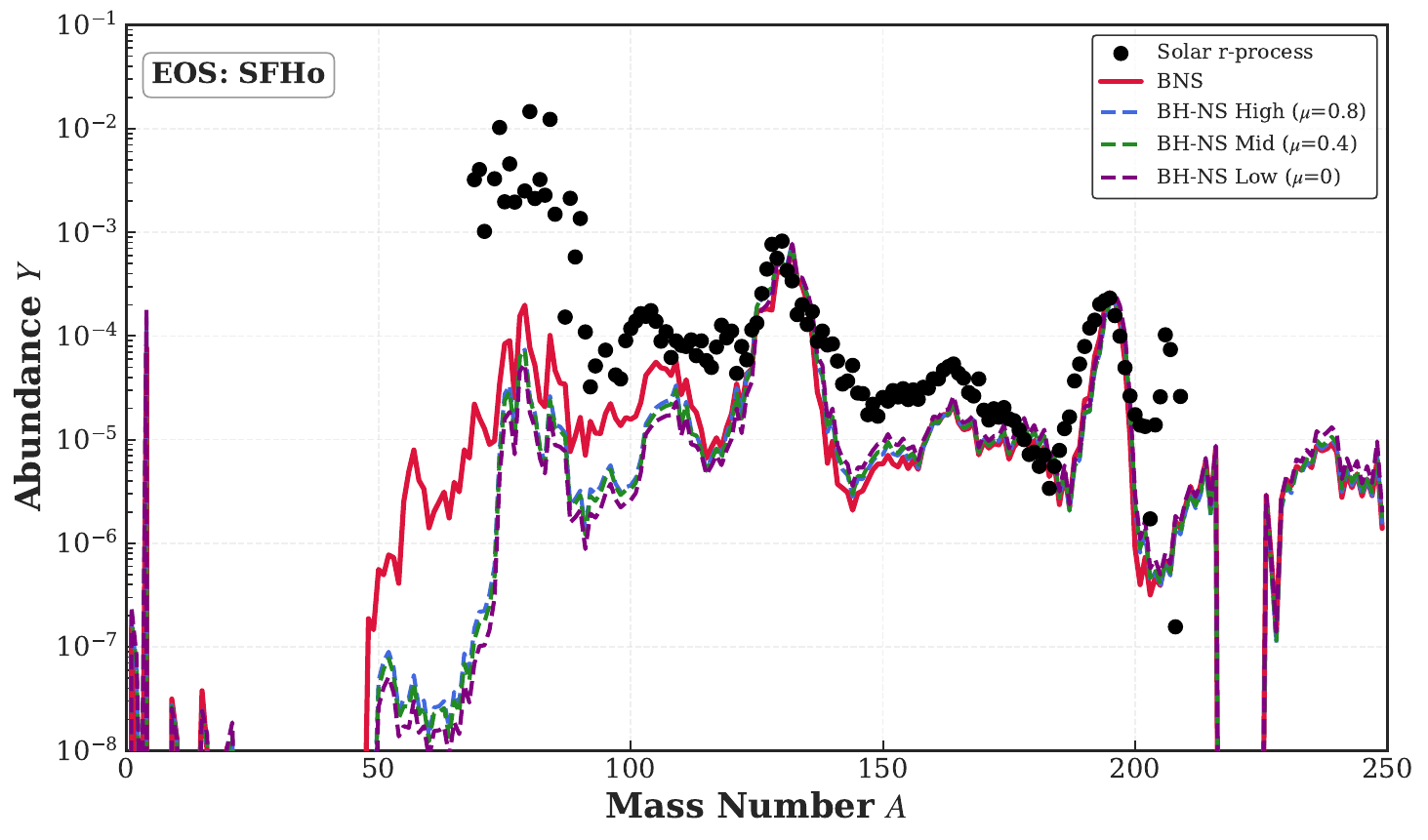}\hfill
\makebox[0.32\linewidth]{}

\caption{Comparisons of the simulated $r$-process elemental abundance patterns from BNS and BH-NS mergers with the solar system $r$-process residuals (black dots; \citealt{2007PhR...450...97A}).}
\label{fig:5}
\end{figure}

\begin{figure}[ht!]
\includegraphics[width=0.48\linewidth]{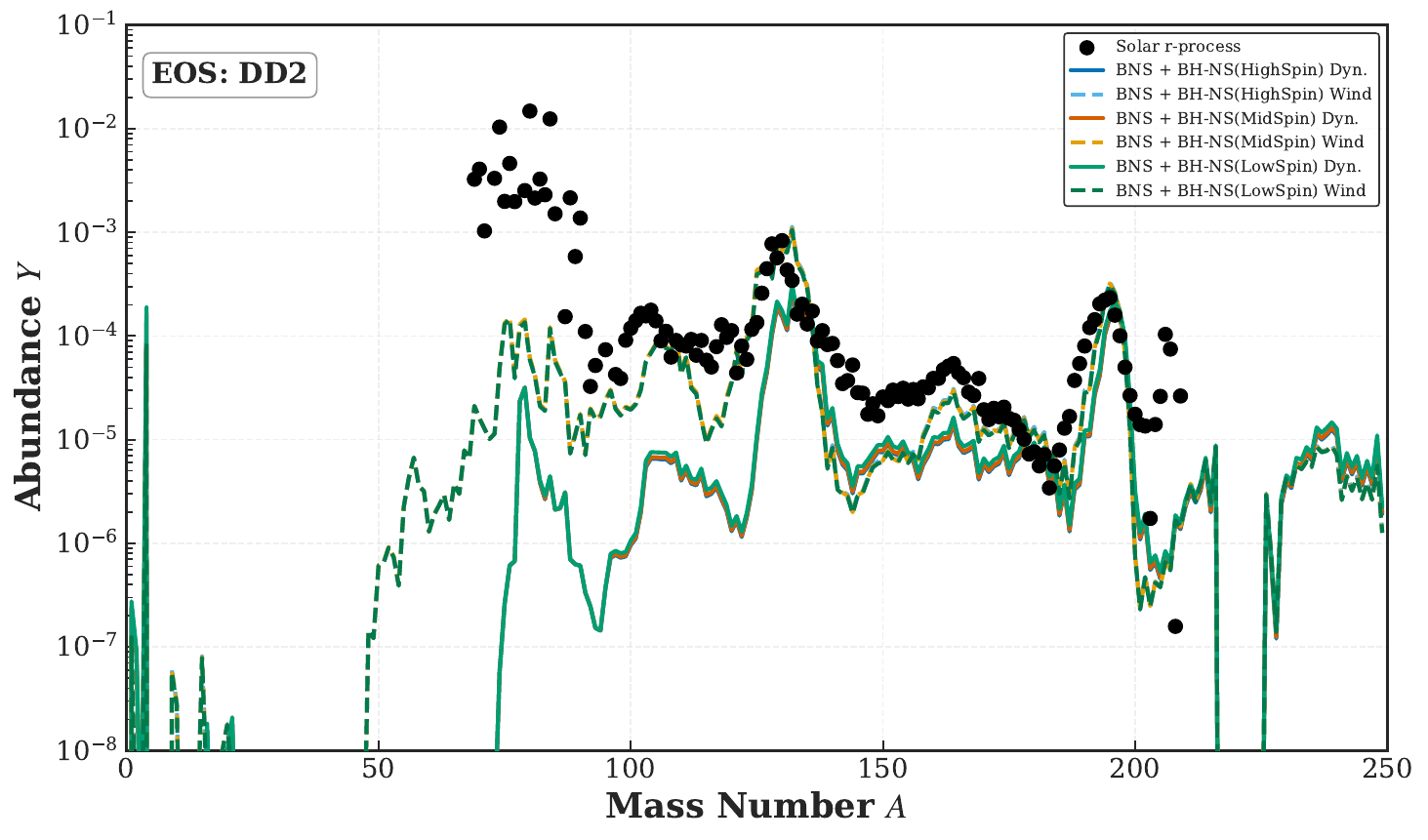}\hfill
\includegraphics[width=0.48\linewidth]{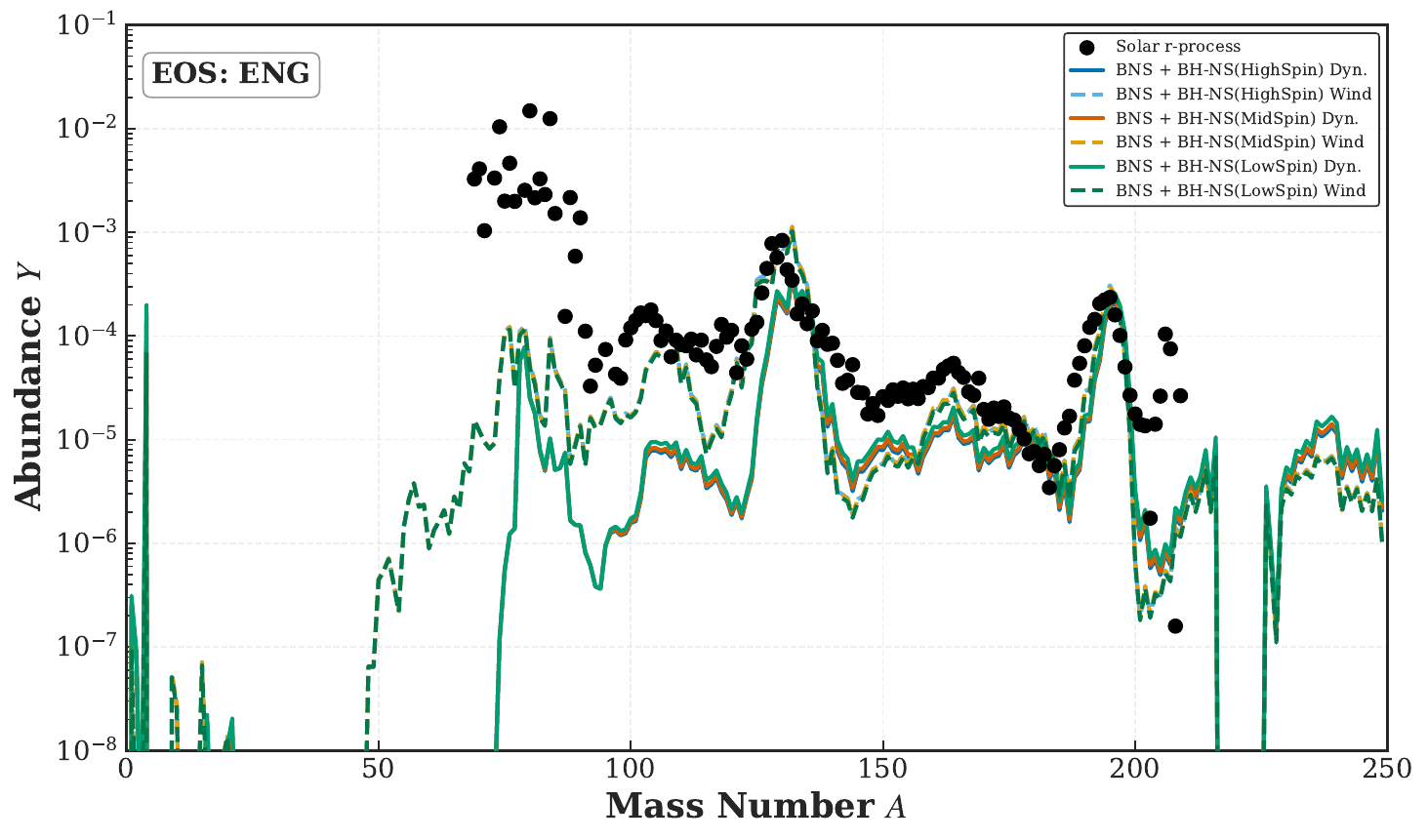}\hfill
\includegraphics[width=0.48\linewidth]{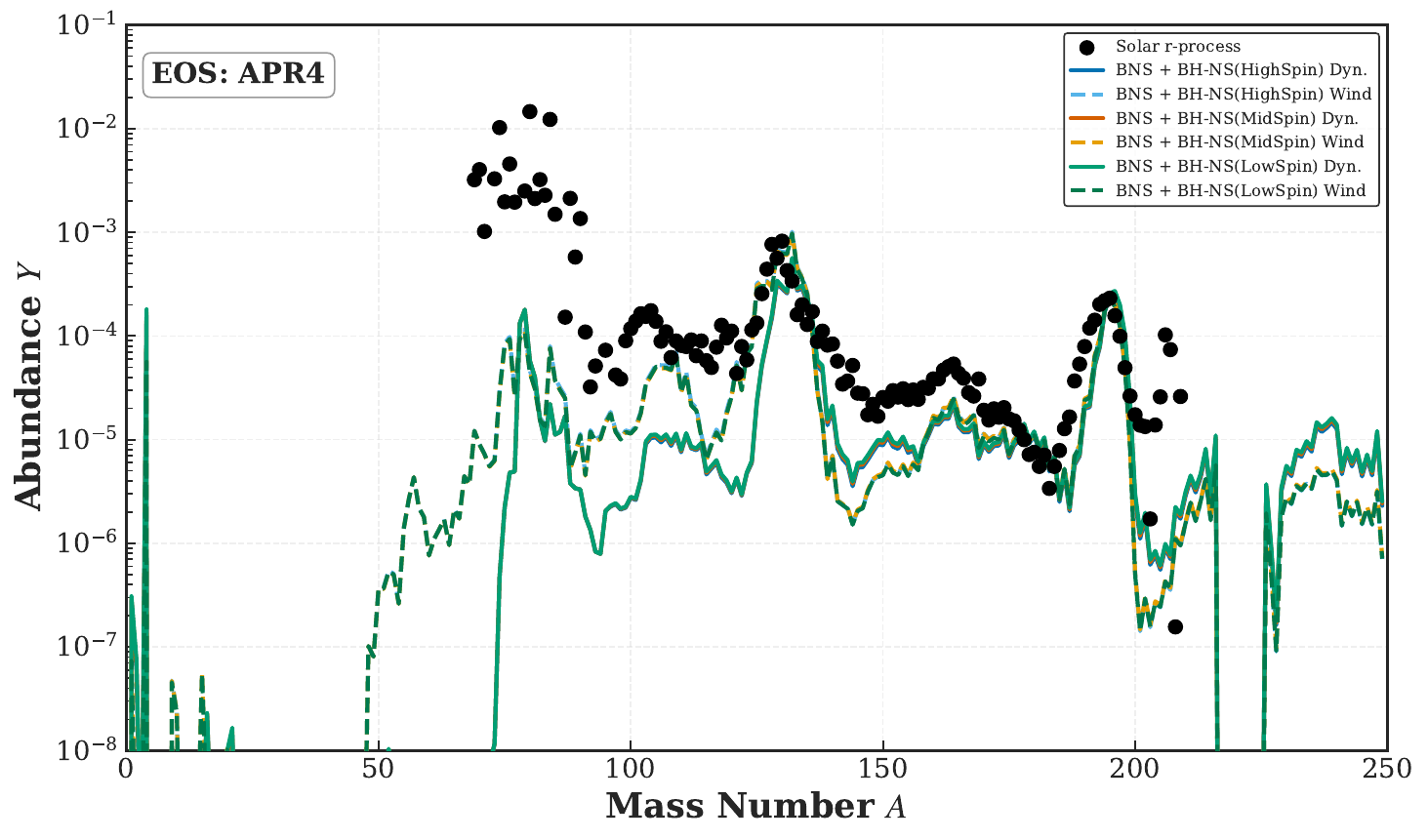}
\includegraphics[width=0.48\linewidth]{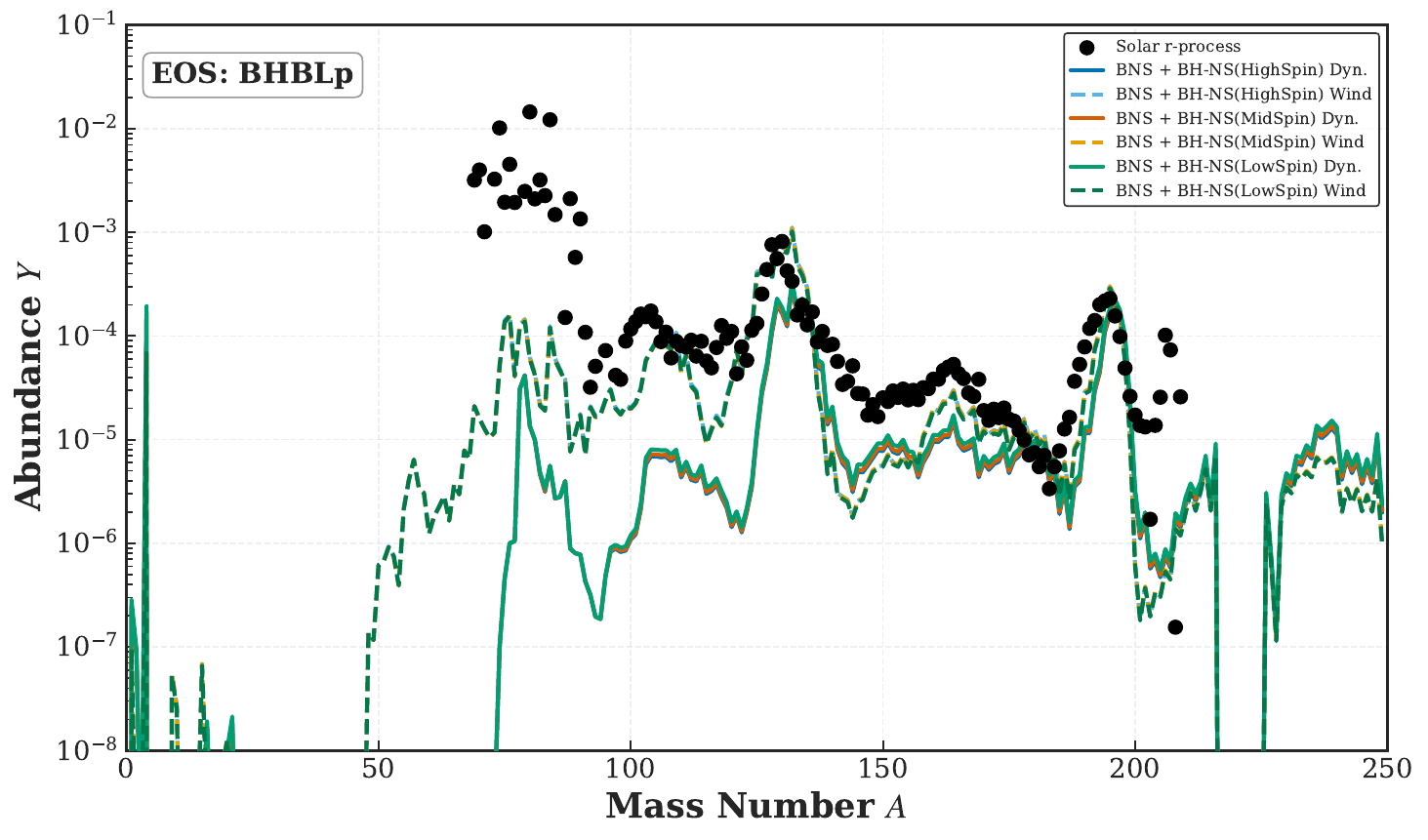}\hfill
\includegraphics[width=0.48\linewidth]{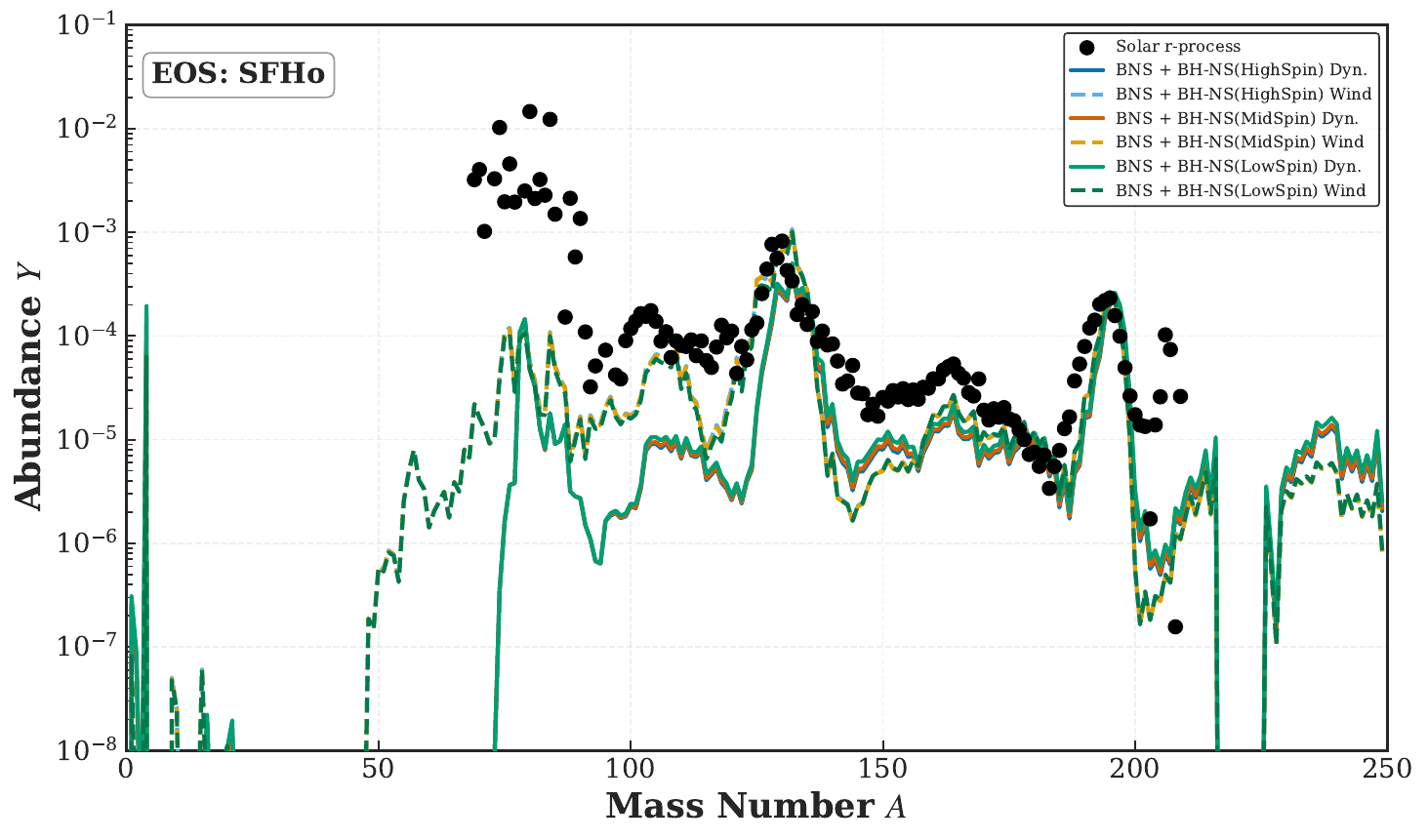}\hfill
\makebox[0.32\linewidth]{}

\caption{Similar to Figure~\ref{fig:5}, Comparison of the simulated $r$-process elemental abundance patterns from dynamical and disk wind ejecta with the solar system $r$-process residuals (black dots; \citealt{2007PhR...450...97A}).}
\label{fig:6}
\end{figure}

The averaged abundance, on the other hand, may be affected by the selected EOS of the NS for BNS systems. Figure~\ref{fig:4} shows the $r$-process event-averaged abundances for five different EOSs of NS for BNS mergers. It is found that the averaged abundance seems not to be significantly dependent on the EOS of NS, but the EOS can affect the total amount of ejecta produced per event, through the maximum mass $M_{\mathrm{max}}$ and compactness $C_{\rm NS}$ of NS  \citep{2020PhRvD.102j3015G,2021PhRvL.126y1101L,2023MNRAS.520.2806C}. For a given EOS of NS, one basic question is what is the contribution of BNS and BH-NS mergers to the production of $r$-process elements. Figure~\ref{fig:5} shows the $r$-process abundance of BNS and BH-NS with different spins of the BH for five different EOSs of the NS, and it is found that the abundance of BNS is higher than that of BH-NS mergers before the second peak (for mass number $A \sim 130$), but it is lower when the mass number is larger than 130. Such a result is consistent with the current theoretical prediction, and it suggests that the merger rate of BNS must be larger than that of BH-NS mergers to produce a larger total ejecta mass if BNS dominates the contribution to production of heavy $r$-process elements. The following question is what is the contribution of the total mass of dynamical ejecta and disk wind to BNS $+$ BH-NS. Similar to Figure~\ref{fig:5}, Figure~\ref{fig:6} shows the $r$-process abundance of the total mass of dynamical ejecta and disk wind by combining BNS and BH-NS mergers for five different EOSs of NS. It is found that the lighter $r$-process elements ($A \lesssim 130$) are mainly produced by the disk wind, while the heavier elements ($A \gtrsim 130$) originate mainly from the dynamical ejecta. Furthermore, we find that the BH spin is not significant enough to affect the heavy element abundance pattern. Such a result is also in line with the physically expected result, because the spin is primarily a dynamical parameter that does not dictate the nucleosynthetic yields.

\section{Contribution to the Galactic Lanthanides from Compact Object Mergers} \label{sec:gce}
In order to compare the contribution of heavier elements from BNS and BH-NS mergers with the distribution of observed lanthanides like europium (Eu) in our Galaxy, we investigate the cumulative effects of BNS and BH-NS mergers over cosmic time by establishing the nucleosynthetic properties with the GCE model. Due to the significant dispersion of observational data for Eu in the Milky Way, we adopt a stochastic multi-zone model instead of a single-zone of GCE model \citep{2004A&A...416..997A}. Within this scenario, the star-forming gas is divided into independent cells. A large number of supernovae (CCSNe and SNe Ia) provide a smooth Fe enrichment, while rare compact binary mergers (e.g., BNS and BH-NS) can result in a highly stochastic Eu injection across different zones. We set the zone mass as $M_{\mathrm{zone}}=3 \times 10^5\,M_\odot$, which represents the typical gas mass mixed by a single supernova remnant \citep{1998ApJ...500...95T}. The total mass of Galactic gas is about $M_{\mathrm{gas}}\sim 10^{10}\,M_\odot$ \citep{2006PASJ...58..847N}, and the number of zones can be estimated as $N_{\mathrm{zone}}^{\mathrm{gal}} = M_{\mathrm{gas}} / M_{\mathrm{zone}} \approx 3.3 \times 10^4$ zones. The global star formation history (SFH) is evolved with a delayed exponential model from \citet{1986A&A...161...89S} and \citet{2001AJ....121..753B},
\begin{equation}
\psi(t) = 25.0 \left(\frac{t}{t_0}\right) \exp\left(-\frac{t}{t_0}\right) \,M_\odot\,\mathrm{yr}^{-1},
\label{eq:sfr}
\end{equation}
where $t_0 = 2.5\,\mathrm{Gyr}$ characterizes the peak time of star formation, and the normalization constant of $25.0\,M_\odot\,\mathrm{yr}^{-1}$ ensures a total stellar mass which is consistent with the estimated mass of the Milky Way.

The global merger rate $R_{\mathrm{merger}}(\tau)$ at cosmic time $\tau$ can be calculated by convolving the SFH with the normalized delay-time distribution, denoted as $\Phi(t) \propto t^{-1.1}$, over the delay time $t$:
\begin{equation}
R_{\mathrm{merger}}(\tau) = \eta \int_{0.035}^{\tau} \psi(\tau-t) \, \Phi(t) \, dt,
\label{eq:merger_rate}
\end{equation}
where $\eta$ is the merger efficiency on cosmological scales, and it represents the number of mergers per unit stellar mass formed in units of $M_\odot^{-1}$. The efficiency $\eta$ is calibrated against the local volumetric merger rate $\mathcal{R}$ from the LIGO GWTC-3 catalog \citep{2023PhRvX..13a1048A} by
\begin{equation}
\eta = \frac{\mathcal{R}}{\rho_{\mathrm{SFR}}}.
\label{eq:eta_conversion} 
\end{equation}
Here, $\rho_{\mathrm{SFR}}$ is the density of star formation rate. One needs to note that the total BNS merger rate is about $(28-300)\,\mathrm{Gpc}^{-3}\,\mathrm{yr}^{-1}$ to analyze GWTC-4 \citep{2026ApJ..1005L..51A, 2026ApJ..1006L..49F}. The cosmological short GRB rate, $r$-process rate, and rate inferred from Galactic BNSs, are a factor of several to several dozen higher than the BNS rate estimated from GWTC-4 \citep{2026ApJ..1006L..49F}. By adopting $\rho_{\mathrm{SFR}} \approx 1.5 \times 10^{7}\,M_\odot\,\mathrm{Gpc}^{-3}\,\mathrm{yr}^{-1}$ in a local universe \citep{2014ARA&A..52..415M}, the observational constraints $\mathcal{R}_{\mathrm{BNS}} \sim 350\,\mathrm{Gpc}^{-3}\,\mathrm{yr}^{-1}$, and $\mathcal{R}_{\mathrm{BH-NS}}=(7.8$--$140)\,\mathrm{Gpc}^{-3}\,\mathrm{yr}^{-1}$ \citep{2023PhRvX..13a1048A}, we derive typical efficiencies of $\eta_{\mathrm{BNS}} \approx 2.3\times10^{-5}\,M_\odot^{-1}$ for BNS, and $\eta_{\mathrm{BH-NS}} \sim 5.2\times10^{-7}\,M_\odot^{-1}$, $1\times10^{-6}\,M_\odot^{-1}$, $6\times10^{-6}\,M_\odot^{-1}$ for the high, medium, and low spin, respectively for BH-NS mergers \footnote{One needs to state that we adopt the data of BHs with measured the spin from GWTC-3 events \citep{2024PhRvL.133e1401L}. The faction of number for BHs with high-spin, mid-spin, and low-spin are roughly to be 7:13:80, and we adopt $\mathcal{R}_{\mathrm{BH-NS}}=7.8, 15, 90\,\mathrm{Gpc}^{-3}\,\mathrm{yr}^{-1}$ for high-spin, mid-spin, and low-spin, respectively for BH-NS mergers}. 

\begin{figure}[ht!]
\centering
\includegraphics[width=0.7\linewidth]{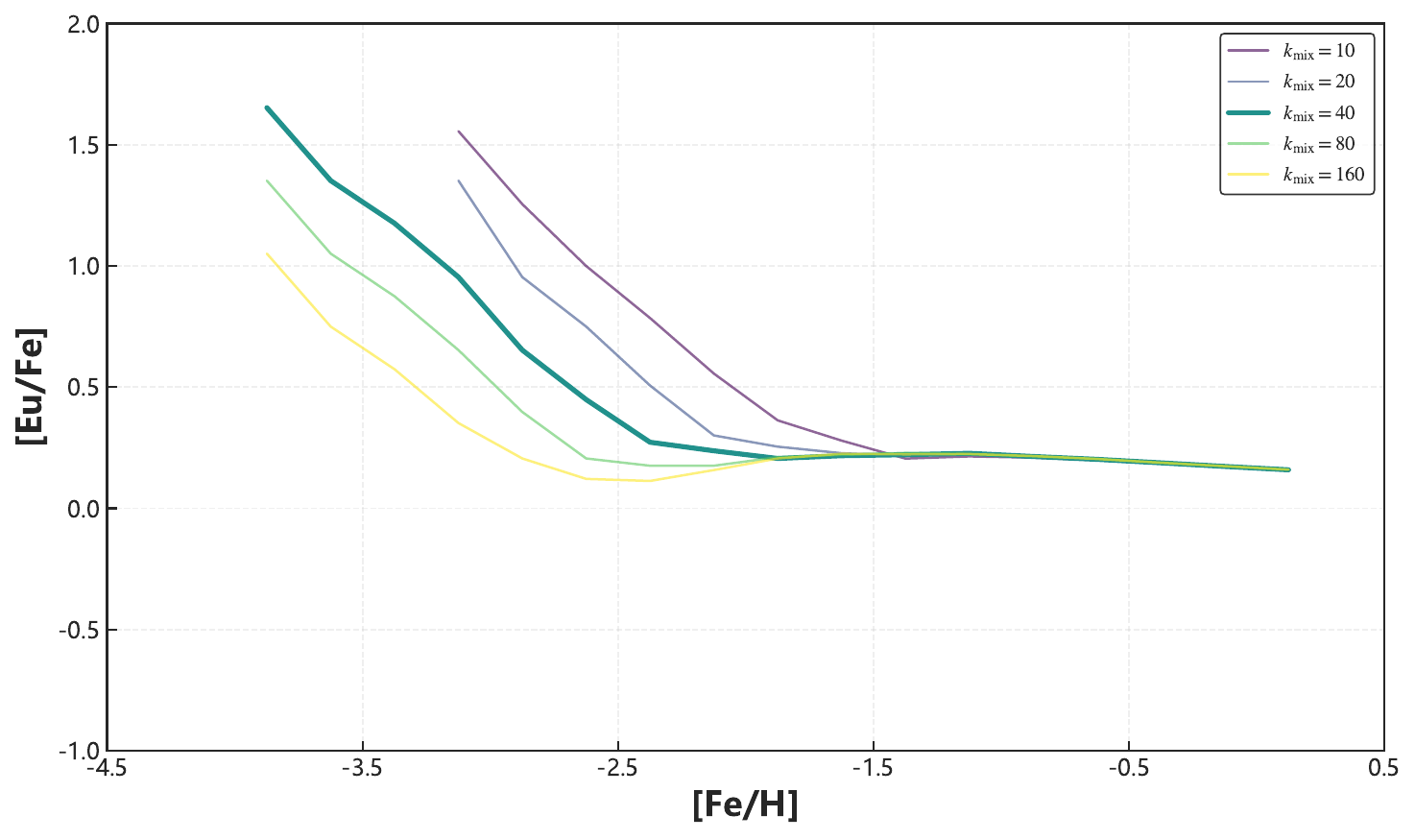}
\caption{[Eu/Fe] as a function of [Fe/H] for adopting $k_{\mathrm{mix}}=10$, 20, 40, 80, and 160.}
\label{fig:7}
\end{figure}

Furthermore, the heavy elements produced by the $r$-process from ejecta with a velocity of (0.1-0.2)$c$ for BNS and BH-NS mergers, which not only affects the abundance of heavy elements in the local place, but also influences the abundance of heavy elements in the surrounding region. Physically, BNS and BH-NS mergers result in a blast wave which sweeps up the surrounding interstellar medium (ISM), thereby mixing and diluting any heavy element enhancement supplied by the ejecta. Therefore, we introduce a mixing factor ($k_{\mathrm{mix}}$) which is used to describe how large the diffusion of these ejecta is. The definition of $k_{\mathrm{mix}}$ is the number of contiguous spatial cells across by the nucleosynthetic products of a single merger event are evenly distributed. In order to test how sensitive the $k_{\mathrm{mix}}$ value is to abundance dispersion, we adopt $k_{\mathrm{mix}}=10$, 20, 40, 80, and 160 to plot in the $\rm Eu/Fe-Fe/H$ diagram (see Figure~\ref{fig:7}). It is found that the values of $k_{\mathrm{mix}}$ in the early stage of GCE have a significant impact on the dispersion of heavy element abundances, but the effect is insensitive in the later stage of GCE. Therefore, we set $k_{\mathrm{mix}}=40$ which effectively corresponds to a total mixing mass of $M_{\mathrm{mix}} = k_{\mathrm{mix}} M_{\mathrm{zone}} \approx 10^7\,M_\odot$ \citep{2016ApJ...832..149B, 2018ApJ...860...89M}. This amplifies the effective Poisson sampling rate to $\lambda_{\mathrm{eff}} = R_{\mathrm{merger}}(t) \, \Delta t \cdot k_{\mathrm{mix}} / N_{\mathrm{zone}}^{\mathrm{gal}}$, effectively spreading the event across adjacent gas zones $\sim 40$.

CCSNe and SNe Ia, on the other hand, can also produce fixed Fe yields of $0.058\,M_\odot$ and $0.7\,M_\odot$ per event, respectively \citep{2008MNRAS.384..267M, 2024ApJ...973..122W}. Together with the high occurrence rates of both CCSNe ($\eta\approx 0.01\,M_\odot^{-1}$) \citep{2001MNRAS.322..231K} and SNe Ia ($\eta\approx 0.002\,M_\odot^{-1}$) \citep{2014ARA&A..52..107M}, the Poisson noise is heavily suppressed which can result in Fe enrichment to proceed in a statistically continuous manner. For the heavy element Eu in compact binary mergers, we dynamically assign the Eu ejecta mass for each merger event by randomly sampling from our simulated binary population database. By considering the discrepancy between theoretical nucleosynthesis yields and observations in the lanthanide region, the systematic uncertainties around $A \sim 150-160$ introduced by fission fragment distribution assumptions, and the variations arising from the choice of different nuclear mass models, can result in a difference of a factor of 1 to 3 between theoretical nucleosynthesis yields and observations \citep{2021RvMP...93a5002C,2024MNRAS.529.1154C, 2025A&A...694A.180M}. Here, we adopt a conservative scaling factor of $2.0$ in the calculations of the final Eu yields. Also, the extreme rarity of compact binary mergers can cause Eu enrichment to manifest as a pronounced abundance dispersion. Such dichotomy in enrichment behavior arises naturally from the drastically distinct occurrence rates of these source populations. 

\begin{figure}[ht!]
\includegraphics[width=0.48\linewidth]{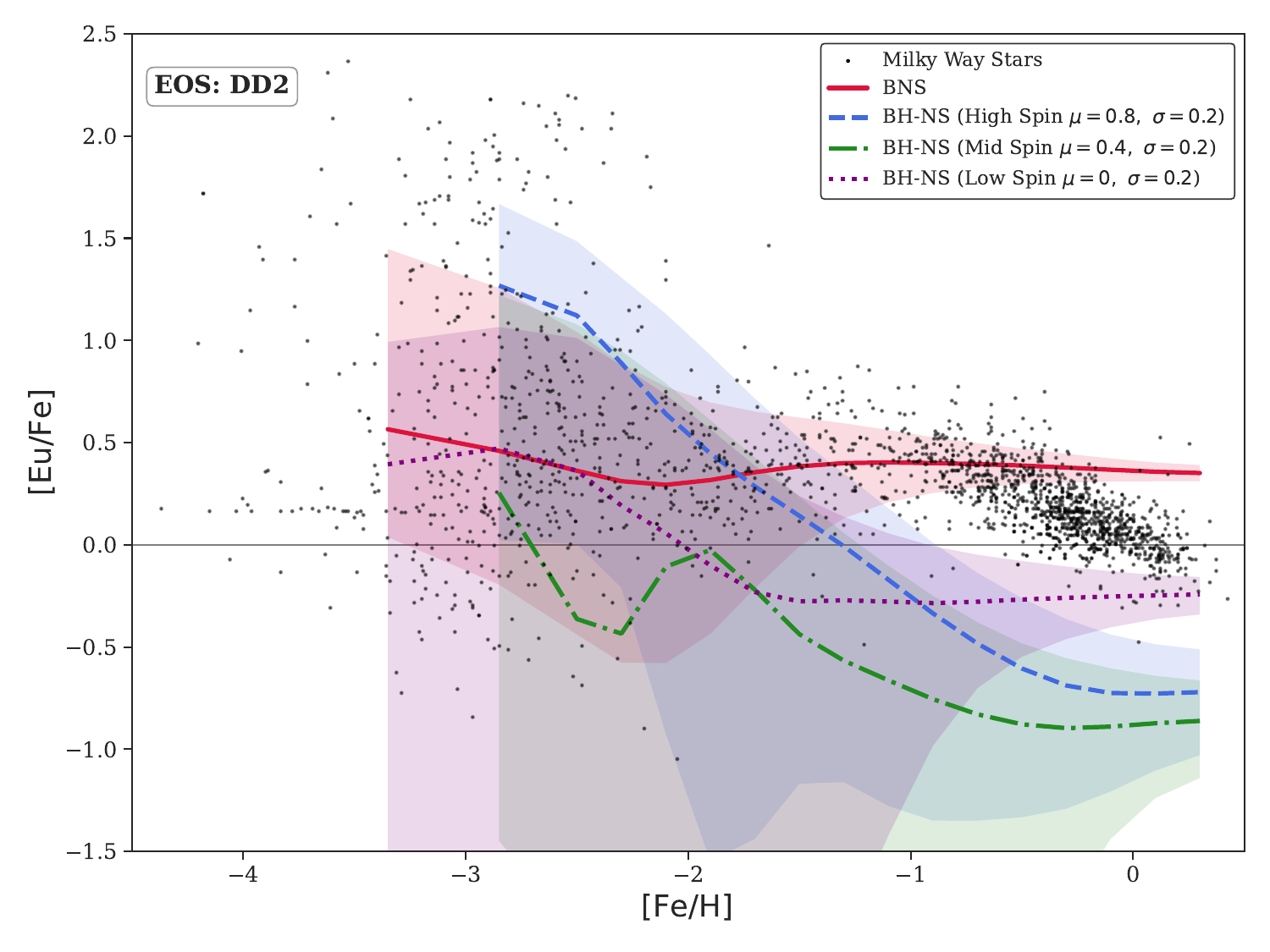}\hfill
\includegraphics[width=0.48\linewidth]{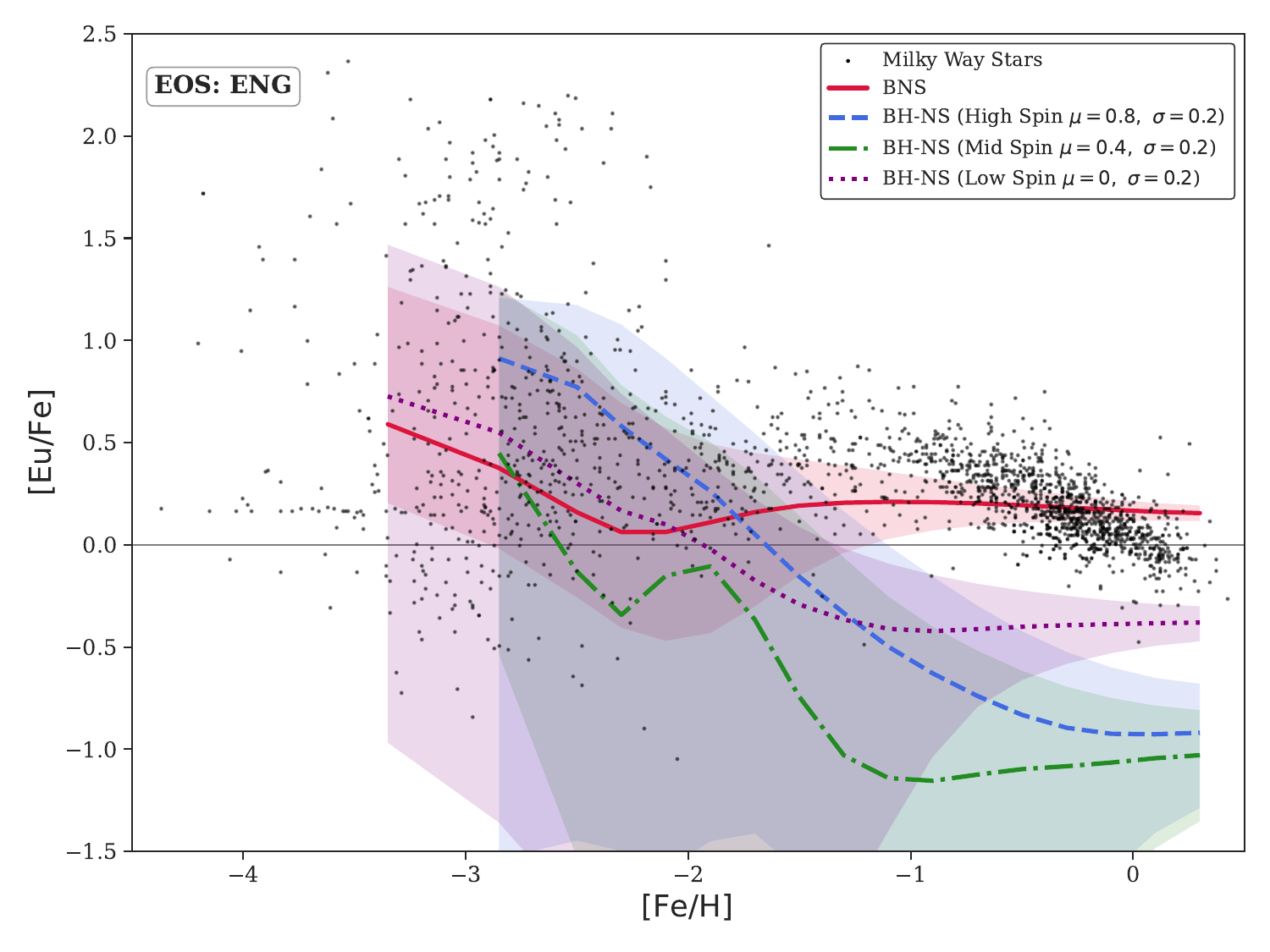}\hfill
\includegraphics[width=0.48\linewidth]{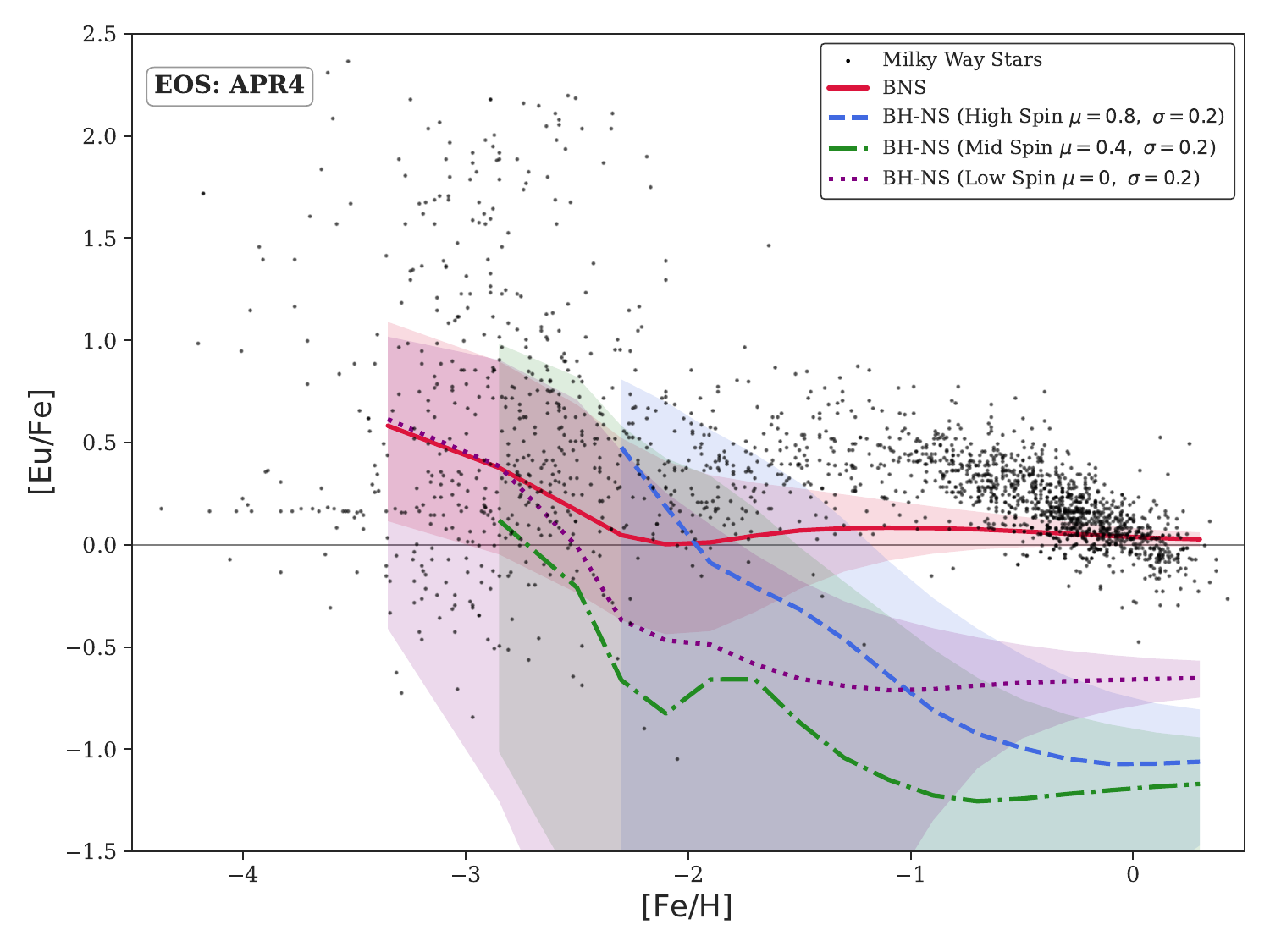}
\includegraphics[width=0.48\linewidth]{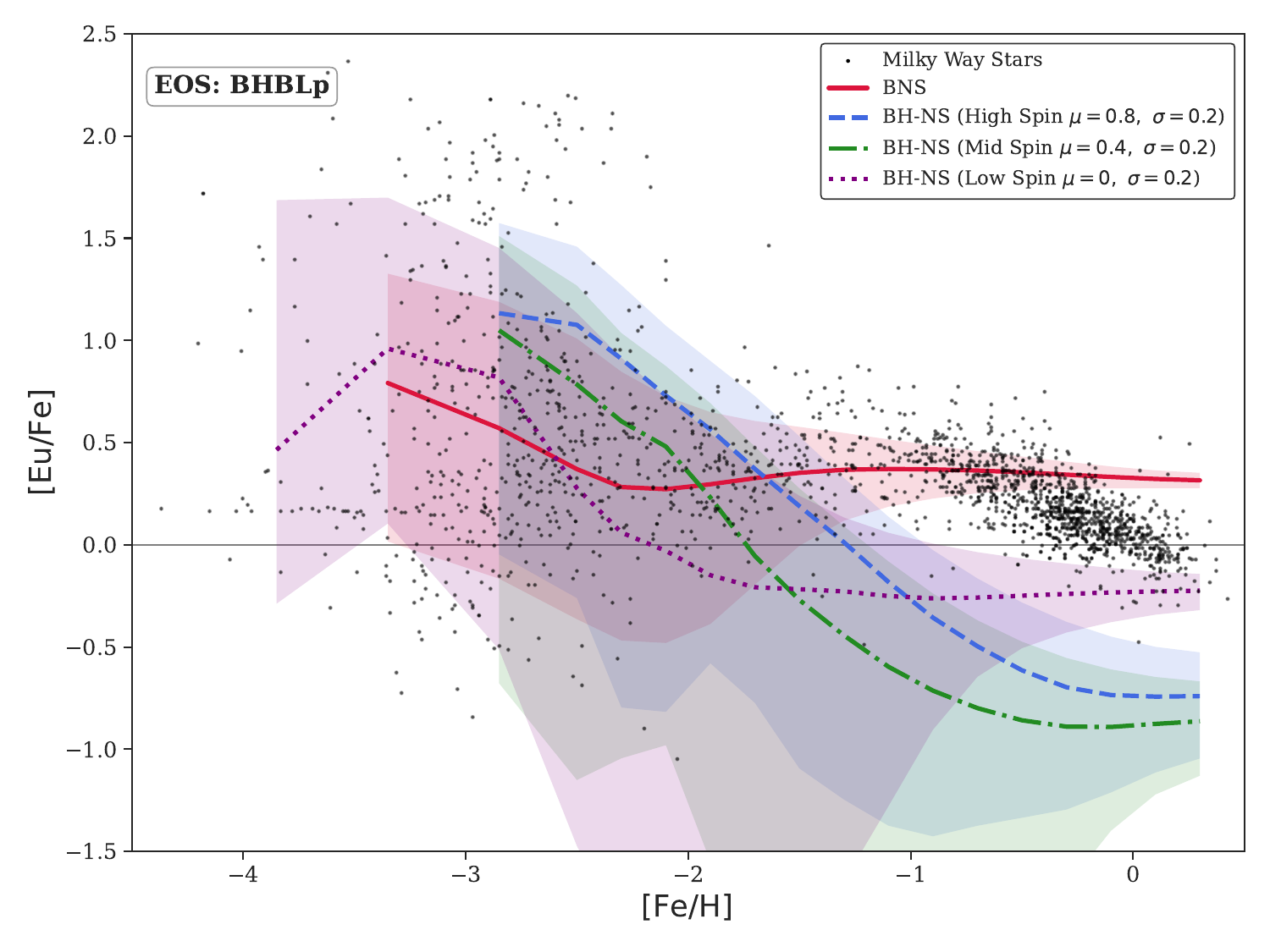}\hfill
\includegraphics[width=0.48\linewidth]{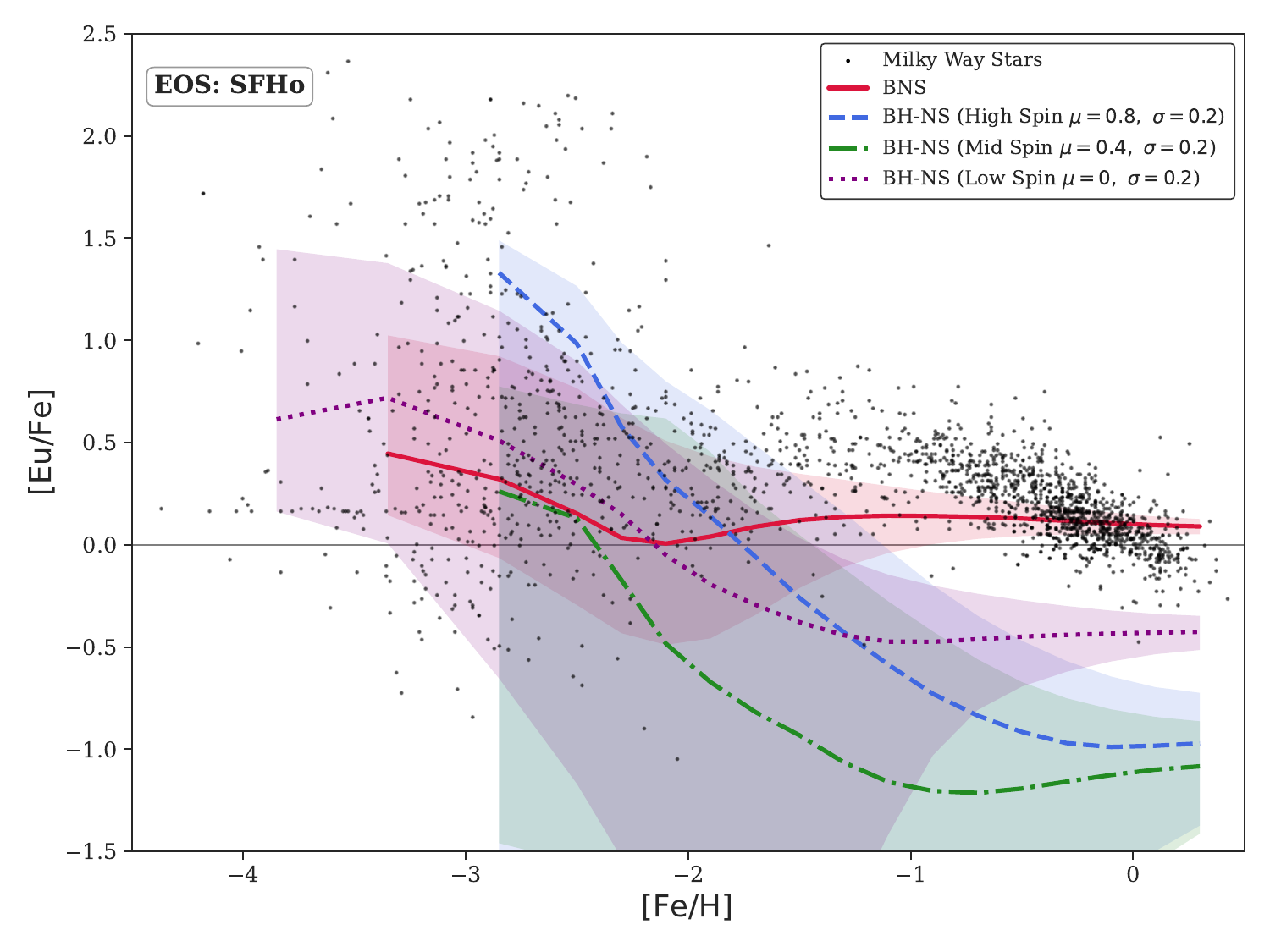}\hfill
\makebox[0.32\linewidth]{}
\caption{$[\mathrm{Eu/Fe}]$ as a function of $[\mathrm{Fe/H}]$. The plot displays the evolutionary tracks of both BNS and BH-NS mergers with different BH spin for five different EOSs. The central tracks and corresponding shaded regions are the median 
evolutionary trajectories and their $1\sigma$ abundance dispersion intervals for BNS (red), high-spin BH-NS (blue), midspin BH-NS (green), and low-spin BH-NS (purple), respectively.} The observational data for $[\mathrm{Eu/Fe}]$ in Milky Way stars are sourced from the database compiled by NuPyCEE's STELLAB module \citep{2017ApJ...835..128C}, which includes data from \citet{2009ApJ...698.1963R, 2014AJ....147..136R, 2015ApJ...807..171J, 2004AJ....128.1177V, 2016A&A...586A..49B}.
\label{fig:8}
\end{figure}

Within this scenario, Figure~\ref{fig:8} shows the evolutionary tracks of the models (e.g., the BNS merger and the BH-NS mergers with three different BH spins) in the $\rm Eu/Fe-Fe/H$ diagram. The central tracks and corresponding shaded regions are the median 
evolutionary trajectories and their $1\sigma$ shaded regions for BNS (red), high-spin BH-NS (blue), mid-spin BH-NS (green), and low-spin BH-NS (purple), respectively. It is found that the constraints from purely BNS mergers are distributed in a relatively narrow abundance dispersion, and it should be the dominant contributor to the production of heavy $r$-process elements at the later stage of GCE. However, for BH-NS mergers, the event-to-event variance in Eu yields becomes extremely large with different spins of the BH. It can result in a substantial abundance dispersion at the early stage of GCE, but has little contribution to the production of heavy $r$-process elements at the later stage of GCE.

\section{Conclusions and Discussion} \label{sec:conclusion}

The multi-messenger observations of GW170817/GRB 170817A/AT2017gfo have unambiguously confirmed that BNS mergers are one of the major astrophysical sites for the $r$-process to produce heavier elements \citep{2017PhRvL.119p1101A, 2017ApJ...848L..17C, 2017ApJ...848L..18N, 2021ApJ...919...59C}. BH-NS systems are another possible astrophysical site for the $r$-process to produce heavier elements with a lower contribution efficiency \citep{2021ARNPS..71..433L,2021LRR....24....5K}. However, several questions naturally arise. What is the contribution of BNS and BH-NS mergers to the production of $r$-process elements? What is the contribution of the total mass of dynamical ejecta and disk wind to BNS$+$BH-NS? Whether BNS mergers alone can self-consistently explain the abundance dispersion of GCE in early MPS?

In this paper, by considering five EOSs of NS and different spins (e.g., high, medium, and low spin) of BHs in BNS and BH-NS mergers, we present a comprehensive framework to couple COMPAS binary population synthesis, nuclear reaction networks (SkyNet nucleosynthesis), and a stochastic GCE model to diagnose the above questions. By adopting the initial binary star parameters and environmental metallicity from COMPAS \citep{2022ApJS..258...34R}, we initialize $10^6$ zero-age main sequence binary star systems, and obtain the total number of SNe ($N=$558880), BNS ($N=$188), and BH-NS ($N=$1436). The following interesting results are obtained.
\begin{itemize}
 \item Most SNe ($\sim 97.3\%$) are concentrated at $<50\,\mathrm{Myr}$, and about $47.3\%$ (or $26.0\%$) of BNS (or BH-NS) are distributed $<100\,\mathrm{Myr}$. The higher spin of BH in BH-NS merger systems corresponds to a larger dynamical ejecta mass.
 \item The averaged abundance seems not to be significantly dependent on the EOS of NS. For the given EOSs of NS, the abundance of BNS is higher than that of BH-NS mergers before the second peak (for mass number $A\sim$130), but it is lower when the mass number is greater than 130.
 \item The lighter $r$-process elements ($A<$130) are mainly produced by the disk wind, while the heavier elements ($A>$130) originate mainly from the dynamical ejecta. The BH spin does not significantly affect the heavy element abundance pattern for BH-NS mergers.
 \item In the $\rm Eu/Fe-Fe/H$ diagram of evolutionary tracks, the constraints from purely BNS mergers are distributed in a relatively narrow abundance dispersion, and it should be the dominant contributor to the production of heavy $r$-process elements at the later stage of GCE. However, for BH-NS mergers, different spins of BH can result in a substantial abundance dispersion at the early stage of GCE, but has a little contributions to the production of heavy $r$-process elements at the later stage of GCE.
\end{itemize}

BNS and BH-NS mergers can indeed provide evidence for the origin of heavier $r$-process elements, especially for those elements with $A>130$ in the Milky Way. However, one needs to note that the contributions to the abundances for heavy $r$-process elements from BNS and BH-NS mergers are significantly lower than that of observational data of solar system when the mass number $A$ is less than 130 (second peak). It suggests that the CCSNe may make a more significant contribution compared with BNS and BH-NS mergers for those elements with lower mass number \citep{2004ApJ...601..864T,2019ApJ...885...33H,2019Natur.569..241S}. Moreover, magnetar giant flares (GFs) have been proposed as other possible sites for $r-$process nucleosynthesis \citep{2024MNRAS.528.5323C}, but the contribution from GFs focuses on the light $r$-process elements due to their lower ejecta mass \citep{2025ApJ...985..234P,2026SCPMA..6909513Q,2026A&A...713A.307Z}.

Recently, \cite{2026ApJ...997L..14W} proposed that an alternative site for lanthanide production is robustly proton-rich matter outflows from CCSNe in the environment with high-entropy winds. After the charged-particle reactions stopped, the excess protons can continue to be converted to neutrons via (anti) neutrino interactions to produce a neutron-capture reaction flow up to $A\sim$ 200, such a process is named the "$\nu i$ process" \citep{2026ApJ...997L..14W}. Based on the results in Section 4, although purely BNS mergers seem to be the dominant contributor to the production of heavy $r$-process elements at a later stage of GCE, their contribution is still lower than that of observational data. It means that the contributions from $\nu i$-process of CCSNe cannot be ignored.

\begin{acknowledgments}
This work is supported by the Natural Science Foundation of China (grant Nos. 12494574, 12403043, and 12621007), the Natural Science Foundation of Guangxi (grant No. 2025GXNSFDA02850010), the Program of Bagui Scholars Program (LHJ), the Guangxi Talent Program (Highland of Innovation Talents), and Guangxi Science and Technology Innovation Platform Program (Leitai Action Plan) (Grant No. Guike LT2600640026 ).
\end{acknowledgments}

\bibliographystyle{aasjournalv7}
\bibliography{ms}

@article{2026ApJ..1005L..51A,
    author          = {{Abac}, A.~G. and {Abouelfettouh}, I. and {Acernese}, F. and {Ackley}, K. and {Adamcewicz}, C. and {Adhicary}, S. and {Adhikari}, D. and {Adhikari}, N. and {Adhikari}, R.~X. and {Adkins}, V.~K. and {Afroz}, S. and {Agarwal}, D. and {Agathos}, M. and {Aghaei Abchouyeh}, M. and {Aguiar}, O.~D. and {Ahmadzadeh}, S. and {Aiello}, L. and {Ain}, A. and {Ajith}, P. and {Akutsu}, T. and {Albanesi}, S. and {Alfaidi}, R.~A. and {Al-Jodah}, A. and {All{\'e}n{\'e}}, C. and {Allocca}, A. and {Al-Shammari}, S. and {Altin}, P.~A. and {Alvarez-Lopez}, S. and {Amarasinghe}, O. and {Amato}, A. and {Amra}, C. and {Ananyeva}, A. and {Anderson}, S.~B. and {Anderson}, W.~G. and {Andia}, M. and {Ando}, M. and {Andrade}, T. and {Andr{\'e}s-Carcasona}, M. and {Andri{\'c}}, T. and {Anglin}, J. and {Ansoldi}, S. and {Antelis}, J.~M. and {Antier}, S. and {Aoumi}, M. and {Appavuravther}, E.~Z. and {Appert}, S. and {Apple}, S.~K. and {Arai}, K. and {Araya}, A. and {Araya}, M.~C. and {Arca Sedda}, M. and {Areeda}, J.~S. and {Argianas}, L. and {Aritomi}, N. and {Armato}, F. and {Armstrong}, S. and {Arnaud}, N. and {Arogeti}, M. and {Aronson}, S.~M. and {Arun}, K.~G. and {Ashton}, G. and {Aso}, Y. and {Assiduo}, M. and {Assis de Souza Melo}, S. and {Aston}, S.~M. and {Astone}, P. and {Attadio}, F. and {Aubin}, F. and {Aultoneal}, K. and {Avallone}, G. and {Babak}, S. and {Badaracco}, F. and {Badger}, C. and {Bae}, S. and {Bagnasco}, S. and {Bagui}, E. and {Baiotti}, L. and {Bajpai}, R. and {Baka}, T. and {Baker}, T. and {Ball}, M. and {Ballardin}, G. and {Ballmer}, S.~W. and {Banagiri}, S. and {Banerjee}, B. and {Bankar}, D. and {Baptiste}, T.~M. and {Baral}, P. and {Barayoga}, J.~C. and {Barish}, B.~C. and {Barker}, D. and {Barman}, N. and {Barneo}, P. and {Barone}, F. and {Barr}, B. and {Barsotti}, L. and {Barsuglia}, M. and {Barta}, D. and {Bartoletti}, A.~M. and {Barton}, M.~A. and {Bartos}, I. and {Basak}, S. and {Basalaev}, A. and {Bassiri}, R. and {Basti}, A. and {Bates}, D.~E. and {Bawaj}, M. and {Baxi}, P. and {Bayley}, J.~C. and {Baylor}, A.~C. and {Baynard}, II, P.~A. and {Bazzan}, M. and {Bedakihale}, V.~M. and {Beirnaert}, F. and {Bejger}, M. and {Belardinelli}, D. and {Bell}, A.~S. and {Bellie}, D.~S. and {Bellizzi}, L. and {Beltran-Martinez}, D. and {Benoit}, W. and {Bentara}, I. and {Bentley}, J.~D. and {Ben Yaala}, M. and {Bera}, S. and {Bergamin}, F. and {Berger}, B.~K. and {Bernuzzi}, S. and {Beroiz}, M. and {Berry}, C.~P.~L. and {Bersanetti}, D. and {Bertolini}, A. and {Betzwieser}, J. and {Beveridge}, D. and {Bevilacqua}, G. and {Bevins}, N. and {Bhandare}, R. and {Bhatt}, R. and {Bhattacharjee}, D. and {Bhaumik}, S. and {Bhowmick}, S. and {Biancalana}, V. and {Bianchi}, A. and {Bilenko}, I.~A. and {Billingsley}, G. and {Binetti}, A. and {Bini}, S. and {Binu}, C. and {Birnholtz}, O. and {Biscoveanu}, S. and {Bisht}, A. and {Bitossi}, M. and {Bizouard}, M.-A. and {Blaber}, S. and {Blackburn}, J.~K. and {Blagg}, L.~A. and {Blair}, C.~D. and {Blair}, D.~G. and {Bobba}, F. and {Bode}, N. and {Boileau}, G. and {Boldrini}, M. and {Bolingbroke}, G.~N. and {Bolliand}, A. and {Bonavena}, L.~D. and {Bondarescu}, R. and {Bondu}, F. and {Bonilla}, E. and {Bonilla}, M.~S. and {Bonino}, A. and {Bonnand}, R. and {Booker}, P. and {Borchers}, A. and {Borhanian}, S. and {Boschi}, V. and {Bose}, S. and {Bossilkov}, V. and {Boudon}, A. and {Bozzi}, A. and {Bradaschia}, C. and {Brady}, P.~R. and {Branch}, A. and {Branchesi}, M. and {Braun}, I. and {Briant}, T. and {Brillet}, A. and {Brinkmann}, M. and {Brockill}, P. and {Brockmueller}, E. and {Brooks}, A.~F. and {Brown}, B.~C. and {Brown}, D.~D. and {Brozzetti}, M.~L. and {Brunett}, S. and {Bruno}, G. and {Bruntz}, R. and {Bryant}, J. and {Bu}, Y. and {Bucci}, F. and {Buchanan}, J.},
    title           = "{GWTC-4.0: Population Properties of Merging Compact Binaries}",
    journal         = {\apjl},
    year            = 2026,
    month           = jul,
    volume          = {1005},
    number          = {2},
    eid             = {L51},
    pages           = {L51},
    doi             = {10.3847/2041-8213/ae771e},
    archivePrefix   = {arXiv},
    eprint          = {2508.18083},
    primaryClass    = {astro-ph.HE},
    adsurl          = {https://ui.adsabs.harvard.edu/abs/2026ApJ..1005L..51A}
}

@ARTICLE{2017PhRvL.119p1101A,
       author = {{Abbott}, B.~P. and {Abbott}, R. and {Abbott}, T.~D. and {Acernese}, F. and {Ackley}, K. and {Adams}, C. and {Adams}, T. and {Addesso}, P. and {Adhikari}, R.~X. and {Adya}, V.~B. and {Affeldt}, C. and {Afrough}, M. and {Agarwal}, B. and {Agathos}, M. and {Agatsuma}, K. and {Aggarwal}, N. and {Aguiar}, O.~D. and {Aiello}, L. and {Ain}, A. and {Ajith}, P. and {Allen}, B. and {Allen}, G. and {Allocca}, A. and {Altin}, P.~A. and {Amato}, A. and {Ananyeva}, A. and {Anderson}, S.~B. and {Anderson}, W.~G. and {Angelova}, S.~V. and {Antier}, S. and {Appert}, S. and {Arai}, K. and {Araya}, M.~C. and {Areeda}, J.~S. and {Arnaud}, N. and {Arun}, K.~G. and {Ascenzi}, S. and {Ashton}, G. and {Ast}, M. and {Aston}, S.~M. and {Astone}, P. and {Atallah}, D.~V. and {Aufmuth}, P. and {Aulbert}, C. and {AultONeal}, K. and {Austin}, C. and {Avila-Alvarez}, A. and {Babak}, S. and {Bacon}, P. and {Bader}, M.~K.~M. and {Bae}, S. and {Bailes}, M. and {Baker}, P.~T. and {Baldaccini}, F. and {Ballardin}, G. and {Ballmer}, S.~W. and {Banagiri}, S. and {Barayoga}, J.~C. and {Barclay}, S.~E. and {Barish}, B.~C. and {Barker}, D. and {Barkett}, K. and {Barone}, F. and {Barr}, B. and {Barsotti}, L. and {Barsuglia}, M. and {Barta}, D. and {Barthelmy}, S.~D. and {Bartlett}, J. and {Bartos}, I. and {Bassiri}, R. and {Basti}, A. and {Batch}, J.~C. and {Bawaj}, M. and {Bayley}, J.~C. and {Bazzan}, M. and {B{\'e}csy}, B. and {Beer}, C. and {Bejger}, M. and {Belahcene}, I. and {Bell}, A.~S. and {Berger}, B.~K. and {Bergmann}, G. and {Bernuzzi}, S. and {Bero}, J.~J. and {Berry}, C.~P.~L. and {Bersanetti}, D. and {Bertolini}, A. and {Betzwieser}, J. and {Bhagwat}, S. and {Bhandare}, R. and {Bilenko}, I.~A. and {Billingsley}, G. and {Billman}, C.~R. and {Birch}, J. and {Birney}, R. and {Birnholtz}, O. and {Biscans}, S. and {Biscoveanu}, S. and {Bisht}, A. and {Bitossi}, M. and {Biwer}, C. and {Bizouard}, M.~A. and {Blackburn}, J.~K. and {Blackman}, J. and {Blair}, C.~D. and {Blair}, D.~G. and {Blair}, R.~M. and {Bloemen}, S. and {Bock}, O. and {Bode}, N. and {Boer}, M. and {Bogaert}, G. and {Bohe}, A. and {Bondu}, F. and {Bonilla}, E. and {Bonnand}, R. and {Boom}, B.~A. and {Bork}, R. and {Boschi}, V. and {Bose}, S. and {Bossie}, K. and {Bouffanais}, Y. and {Bozzi}, A. and {Bradaschia}, C. and {Brady}, P.~R. and {Branchesi}, M. and {Brau}, J.~E. and {Briant}, T. and {Brillet}, A. and {Brinkmann}, M. and {Brisson}, V. and {Brockill}, P. and {Broida}, J.~E. and {Brooks}, A.~F. and {Brown}, D.~A. and {Brown}, D.~D. and {Brunett}, S. and {Buchanan}, C.~C. and {Buikema}, A. and {Bulik}, T. and {Bulten}, H.~J. and {Buonanno}, A. and {Buskulic}, D. and {Buy}, C. and {Byer}, R.~L. and {Cabero}, M. and {Cadonati}, L. and {Cagnoli}, G. and {Cahillane}, C. and {Calder{\'o}n Bustillo}, J. and {Callister}, T.~A. and {Calloni}, E. and {Camp}, J.~B. and {Canepa}, M. and {Canizares}, P. and {Cannon}, K.~C. and {Cao}, H. and {Cao}, J. and {Capano}, C.~D. and {Capocasa}, E. and {Carbognani}, F. and {Caride}, S. and {Carney}, M.~F. and {Carullo}, G. and {Casanueva Diaz}, J. and {Casentini}, C. and {Caudill}, S. and {Cavagli{\`a}}, M. and {Cavalier}, F. and {Cavalieri}, R. and {Cella}, G. and {Cepeda}, C.~B. and {Cerd{\'a}-Dur{\'a}n}, P. and {Cerretani}, G. and {Cesarini}, E. and {Chamberlin}, S.~J. and {Chan}, M. and {Chao}, S. and {Charlton}, P. and {Chase}, E. and {Chassande-Mottin}, E. and {Chatterjee}, D. and {Chatziioannou}, K. and {Cheeseboro}, B.~D. and {Chen}, H.~Y. and {Chen}, X. and {Chen}, Y. and {Cheng}, H.-P. and {Chia}, H. and {Chincarini}, A. and {Chiummo}, A. and {Chmiel}, T. and {Cho}, H.~S. and {Cho}, M. and {Chow}, J.~H. and {Christensen}, N. and {Chu}, Q. and {Chua}, A.~J.~K. and {Chua}, S.},
        title = "{GW170817: Observation of Gravitational Waves from a Binary Neutron Star Inspiral}",
      journal = {\prl},
         year = 2017,
        month = oct,
       volume = {119},
       number = {16},
          eid = {161101},
        pages = {161101},
          doi = {10.1103/PhysRevLett.119.161101},
archivePrefix = {arXiv},
       eprint = {1710.05832},
 primaryClass = {gr-qc},
       adsurl = {https://ui.adsabs.harvard.edu/abs/2017PhRvL.119p1101A}
}

@article{2023PhRvX..13a1048A,
    author          = {{Abbott}, R. and {Abbott}, T.~D. and {Acernese}, F. and {Ackley}, K. and {Adams}, C. and {Adhikari}, N. and {Adhikari}, R.~X. and {Adya}, V.~B. and {Affeldt}, C. and {Agarwal}, D. and {Agathos}, M. and {Agatsuma}, K. and {Aggarwal}, N. and {Aguiar}, O.~D. and {Aiello}, L. and {Ain}, A. and {Ajith}, P. and {Akutsu}, T. and {de Alarc{\'o}n}, P.~F. and {Akcay}, S. and {Albanesi}, S. and {Allocca}, A. and {Altin}, P.~A. and {Amato}, A. and {Anand}, C. and {Anand}, S. and {Ananyeva}, A. and {Anderson}, S.~B. and {Anderson}, W.~G. and {Ando}, M. and {Andrade}, T. and {Andres}, N. and {Andri{\'c}}, T. and {Angelova}, S.~V. and {Ansoldi}, S. and {Antelis}, J.~M. and {Antier}, S. and {Antonini}, F. and {Appert}, S. and {Arai}, Koji and {Arai}, Koya and {Arai}, Y. and {Araki}, S. and {Araya}, A. and {Araya}, M.~C. and {Areeda}, J.~S. and {Ar{\`e}ne}, M. and {Aritomi}, N. and {Arnaud}, N. and {Arogeti}, M. and {Aronson}, S.~M. and {Arun}, K.~G. and {Asada}, H. and {Asali}, Y. and {Ashton}, G. and {Aso}, Y. and {Assiduo}, M. and {Aston}, S.~M. and {Astone}, P. and {Aubin}, F. and {Austin}, C. and {Babak}, S. and {Badaracco}, F. and {Bader}, M.~K.~M. and {Badger}, C. and {Bae}, S. and {Bae}, Y. and {Baer}, A.~M. and {Bagnasco}, S. and {Bai}, Y. and {Baiotti}, L. and {Baird}, J. and {Bajpai}, R. and {Ball}, M. and {Ballardin}, G. and {Ballmer}, S.~W. and {Balsamo}, A. and {Baltus}, G. and {Banagiri}, S. and {Bankar}, D. and {Barayoga}, J.~C. and {Barbieri}, C. and {Barish}, B.~C. and {Barker}, D. and {Barneo}, P. and {Barone}, F. and {Barr}, B. and {Barsotti}, L. and {Barsuglia}, M. and {Barta}, D. and {Bartlett}, J. and {Barton}, M.~A. and {Bartos}, I. and {Bassiri}, R. and {Basti}, A. and {Bawaj}, M. and {Bayley}, J.~C. and {Baylor}, A.~C. and {Bazzan}, M. and {B{\'e}csy}, B. and {Bedakihale}, V.~M. and {Bejger}, M. and {Belahcene}, I. and {Benedetto}, V. and {Beniwal}, D. and {Bennett}, T.~F. and {Bentley}, J.~D. and {Benyaala}, M. and {Bergamin}, F. and {Berger}, B.~K. and {Bernuzzi}, S. and {Berry}, C.~P.~L. and {Bersanetti}, D. and {Bertolini}, A. and {Betzwieser}, J. and {Beveridge}, D. and {Bhandare}, R. and {Bhardwaj}, U. and {Bhattacharjee}, D. and {Bhaumik}, S. and {Bilenko}, I.~A. and {Billingsley}, G. and {Bini}, S. and {Birney}, R. and {Birnholtz}, O. and {Biscans}, S. and {Bischi}, M. and {Biscoveanu}, S. and {Bisht}, A. and {Biswas}, B. and {Bitossi}, M. and {Bizouard}, M.-A. and {Blackburn}, J.~K. and {Blair}, C.~D. and {Blair}, D.~G. and {Blair}, R.~M. and {Bobba}, F. and {Bode}, N. and {Boer}, M. and {Bogaert}, G. and {Boldrini}, M. and {Bonavena}, L.~D. and {Bondu}, F. and {Bonilla}, E. and {Bonnand}, R. and {Booker}, P. and {Boom}, B.~A. and {Bork}, R. and {Boschi}, V. and {Bose}, N. and {Bose}, S. and {Bossilkov}, V. and {Boudart}, V. and {Bouffanais}, Y. and {Bozzi}, A. and {Bradaschia}, C. and {Brady}, P.~R. and {Bramley}, A. and {Branch}, A. and {Branchesi}, M. and {Brandt}, J. and {Brau}, J.~E. and {Breschi}, M. and {Briant}, T. and {Briggs}, J.~H. and {Brillet}, A. and {Brinkmann}, M. and {Brockill}, P. and {Brooks}, A.~F. and {Brooks}, J. and {Brown}, D.~D. and {Brunett}, S. and {Bruno}, G. and {Bruntz}, R. and {Bryant}, J. and {Bulik}, T. and {Bulten}, H.~J. and {Buonanno}, A. and {Buscicchio}, R. and {Buskulic}, D. and {Buy}, C. and {Byer}, R.~L. and {Cadonati}, L. and {Cagnoli}, G. and {Cahillane}, C. and {Bustillo}, J. Calder{\'o}n and {Callaghan}, J.~D. and {Callister}, T.~A. and {Calloni}, E. and {Cameron}, J. and {Camp}, J.~B. and {Canepa}, M. and {Canevarolo}, S. and {Cannavacciuolo}, M. and {Cannon}, K.~C. and {Cao}, H. and {Cao}, Z. and {Capocasa}, E. and {Capote}, E. and {Carapella}, G.},
    title           = "{Population of Merging Compact Binaries Inferred Using Gravitational Waves through GWTC-3}",
    journal         = {Physical Review X},
    year            = 2023,
    month           = jan,
    volume          = {13},
    number          = {1},
    eid             = {011048},
    pages           = {011048},
    doi             = {10.1103/PhysRevX.13.011048},
    archivePrefix   = {arXiv},
    eprint          = {2111.03634},
    primaryClass    = {astro-ph.HE},
    adsurl          = {https://ui.adsabs.harvard.edu/abs/2023PhRvX..13a1048A}
}

@article{1998PhRvC..58.1804A,
    author          = {{Akmal}, A. and {Pandharipande}, V.~R. and {Ravenhall}, D.~G.},
    title           = "{Equation of state of nucleon matter and neutron star structure}",
    journal         = {\prc},
    year            = 1998,
    month           = sep,
    volume          = {58},
    number          = {3},
    pages           = {1804-1828},
    doi             = {10.1103/PhysRevC.58.1804},
    archivePrefix   = {arXiv},
    eprint          = {nucl-th/9804027},
    primaryClass    = {nucl-th},
    adsurl          = {https://ui.adsabs.harvard.edu/abs/1998PhRvC..58.1804A}
}

@article{2004A&A...416..997A,
    author          = {{Argast}, D. and {Samland}, M. and {Thielemann}, F.-K. and {Qian}, Y.-Z.},
    title           = "{Neutron star mergers versus core-collapse supernovae as dominant r-process sites in the early Galaxy}",
    journal         = {\aap},
    year            = 2004,
    month           = mar,
    volume          = {416},
    pages           = {997-1011},
    doi             = {10.1051/0004-6361:20034265},
    archivePrefix   = {arXiv},
    eprint          = {astro-ph/0309237},
    primaryClass    = {astro-ph},
    adsurl          = {https://ui.adsabs.harvard.edu/abs/2004A&A...416..997A}
}

@ARTICLE{2007PhR...450...97A,
       author = {{Arnould}, M. and {Goriely}, S. and {Takahashi}, K.},
        title = "{The r-process of stellar nucleosynthesis: Astrophysics and nuclear physics achievements and mysteries}",
      journal = {\physrep},
         year = 2007,
        month = sep,
       volume = {450},
       number = {4-6},
        pages = {97-213},
          doi = {10.1016/j.physrep.2007.06.002},
archivePrefix = {arXiv},
       eprint = {0705.4512},
 primaryClass = {astro-ph},
       adsurl = {https://ui.adsabs.harvard.edu/abs/2007PhR...450...97A}
}

@article{2014ApJS..214...22B,
    author          = {{Banik}, Sarmistha and {Hempel}, Matthias and {Bandyopadhyay}, Debades},
    title           = "{New Hyperon Equations of State for Supernovae and Neutron Stars in Density-dependent Hadron Field Theory}",
    journal         = {\apjs},
    year            = 2014,
    month           = oct,
    volume          = {214},
    number          = {2},
    eid             = {22},
    pages           = {22},
    doi             = {10.1088/0067-0049/214/2/22},
    archivePrefix   = {arXiv},
    eprint          = {1404.6173},
    primaryClass    = {astro-ph.HE},
    adsurl          = {https://ui.adsabs.harvard.edu/abs/2014ApJS..214...22B}
}

@article{1972ApJ...178..347B,
    author          = {{Bardeen}, James M. and {Press}, William H. and {Teukolsky}, Saul A.},
    title           = "{Rotating Black Holes: Locally Nonrotating Frames, Energy Extraction, and Scalar Synchrotron Radiation}",
    journal         = {\apj},
    year            = 1972,
    month           = dec,
    volume          = {178},
    pages           = {347-370},
    doi             = {10.1086/151796},
    adsurl          = {https://ui.adsabs.harvard.edu/abs/1972ApJ...178..347B}
}

@article{2013ApJ...773...78B,
    author = {Bauswein, A. and Baumgarte, T. W. and Janka, H.-T.},
    title = {Prompt Merger Collapse and the Maximum Mass of Neutron Stars},
    journal = {Phys. Rev. Lett.},
    year            = 2013,
    month           = {Sep},
    volume          = {111},
    number          = {1},
    eid             = {78},
    pages           = {131101},
    doi             = {10.1103/PhysRevLett.111.131101},
    archivePrefix   = {arXiv},
    eprint          = {1302.6530},
    primaryClass    = {astro-ph.SR},
    adsurl          = {https://link.aps.org/doi/10.1103/PhysRevLett.111.131101}
}

@article{2021A&A...647A.153B,
    author          = {{Bavera}, Simone S. and {Fragos}, Tassos and {Zevin}, Michael and {Berry}, Christopher P.~L. and {Marchant}, Pablo and {Andrews}, Jeff J. and {Coughlin}, Scott and {Dotter}, Aaron and {Kovlakas}, Konstantinos and {Misra}, Devina and {Serra-Perez}, Juan G. and {Qin}, Ying and {Rocha}, Kyle A. and {Rom{\'a}n-Garza}, Jaime and {Tran}, Nam H. and {Zapartas}, Emmanouil},
    title           = "{The impact of mass-transfer physics on the observable properties of field binary black hole populations}",
    journal         = {\aap},
    year            = 2021,
    month           = mar,
    volume          = {647},
    eid             = {A153},
    pages           = {A153},
    doi             = {10.1051/0004-6361/202039804},
    archivePrefix   = {arXiv},
    eprint          = {2010.16333},
    primaryClass    = {astro-ph.HE},
    adsurl          = {https://ui.adsabs.harvard.edu/abs/2021A&A...647A.153B}
}

@ARTICLE{2019MNRAS.487.4847B,
       author = {{Beniamini}, Paz and {Piran}, Tsvi},
        title = "{The Gravitational waves merger time distribution of binary neutron star systems}",
      journal = {\mnras},
         year = 2019,
        month = aug,
       volume = {487},
       number = {4},
        pages = {4847-4854},
          doi = {10.1093/mnras/stz1589},
archivePrefix = {arXiv},
       eprint = {1903.11614},
 primaryClass = {astro-ph.HE},
       adsurl = {https://ui.adsabs.harvard.edu/abs/2019MNRAS.487.4847B}
}

@article{2021MNRAS.508.5028B,
    author          = {{Broekgaarden}, Floor S. and {Berger}, Edo and {Neijssel}, Coenraad J. and {Vigna-G{\'o}mez}, Alejandro and {Chattopadhyay}, Debatri and {Stevenson}, Simon and {Chruslinska}, Martyna and {Justham}, Stephen and {de Mink}, Selma E. and {Mandel}, Ilya},
    title           = "{Impact of massive binary star and cosmic evolution on gravitational wave observations I: black hole-neutron star mergers}",
    journal         = {\mnras},
    year            = 2021,
    month           = dec,
    volume          = {508},
    number          = {4},
    pages           = {5028-5063},
    doi             = {10.1093/mnras/stab2716},
    archivePrefix   = {arXiv},
    eprint          = {2103.02608},
    primaryClass    = {astro-ph.HE},
    adsurl          = {https://ui.adsabs.harvard.edu/abs/2021MNRAS.508.5028B}
}

@article{2018ApJ...855...99C,
    author          = {{C{\^o}t{\'e}}, Benoit and {Fryer}, Chris L. and {Belczynski}, Krzysztof and {Korobkin}, Oleg and {Chru{\'s}li{\'n}ska}, Martyna and {Vassh}, Nicole and {Mumpower}, Matthew R. and {Lippuner}, Jonas and {Sprouse}, Trevor M. and {Surman}, Rebecca and {Wollaeger}, Ryan},
    title           = "{The Origin of r-process Elements in the Milky Way}",
    journal         = {\apj},
    year            = 2018,
    month           = mar,
    volume          = {855},
    number          = {2},
    eid             = {99},
    pages           = {99},
    doi             = {10.3847/1538-4357/aaad67},
    archivePrefix   = {arXiv},
    eprint          = {1710.05875},
    primaryClass    = {astro-ph.GA},
    adsurl          = {https://ui.adsabs.harvard.edu/abs/2018ApJ...855...99C}
}

@article{2022ApJS..258...34R,
    author          = {{Riley}, Jeff and {Agrawal}, Poojan and {Barrett}, Jim W. and {Boyett}, Kristan N.~K. and {Broekgaarden}, Floor S. and {Chattopadhyay}, Debatri and {Gaebel}, Sebastian M. and {Gittins}, Fabian and {Hirai}, Ryosuke and {Howitt}, George and {Justham}, Stephen and {Khandelwal}, Lokesh and {Kummer}, Floris and {Lau}, Mike Y.~M. and {Mandel}, Ilya and {de Mink}, Selma E. and {Neijssel}, Coenraad and {Riley}, Tim and {van Son}, Lieke and {Stevenson}, Simon and {Vigna-G{\'o}mez}, Alejandro and {Vinciguerra}, Serena and {Wagg}, Tom and {Willcox}, Reinhold and {Team Compas}},
    title           = "{Rapid Stellar and Binary Population Synthesis with COMPAS}",
    journal         = {\apjs},
    year            = 2022,
    month           = feb,
    volume          = {258},
    number          = {2},
    eid             = {34},
    pages           = {34},
    doi             = {10.3847/1538-4365/ac416c},
    archivePrefix   = {arXiv},
    eprint          = {2109.10352},
    primaryClass    = {astro-ph.IM},
    adsurl          = {https://ui.adsabs.harvard.edu/abs/2022ApJS..258...34R}
}

@article{2017ApJ...848L..17C,
    author          = {{Cowperthwaite}, P.~S. and {Berger}, E. and {Villar}, V.~A. and {Metzger}, B.~D. and {Nicholl}, M. and {Chornock}, R. and {Blanchard}, P.~K. and {Fong}, W. and {Margutti}, R. and {Soares-Santos}, M. and {Alexander}, K.~D. and {Allam}, S. and {Annis}, J. and {Brout}, D. and {Brown}, D.~A. and {Butler}, R.~E. and {Chen}, H.-Y. and {Diehl}, H.~T. and {Doctor}, Z. and {Drout}, M.~R. and {Eftekhari}, T. and {Farr}, B. and {Finley}, D.~A. and {Foley}, R.~J. and {Frieman}, J.~A. and {Fryer}, C.~L. and {Garc{\'\i}a-Bellido}, J. and {Gill}, M.~S.~S. and {Guillochon}, J. and {Herner}, K. and {Holz}, D.~E. and {Kasen}, D. and {Kessler}, R. and {Marriner}, J. and {Matheson}, T. and {Neilsen}, Jr., E.~H. and {Quataert}, E. and {Palmese}, A. and {Rest}, A. and {Sako}, M. and {Scolnic}, D.~M. and {Smith}, N. and {Tucker}, D.~L. and {Williams}, P.~K.~G. and {Balbinot}, E. and {Carlin}, J.~L. and {Cook}, E.~R. and {Durret}, F. and {Li}, T.~S. and {Lopes}, P.~A.~A. and {Louren{\c{c}}o}, A.~C.~C. and {Marshall}, J.~L. and {Medina}, G.~E. and {Muir}, J. and {Mu{\~n}oz}, R.~R. and {Sauseda}, M. and {Schlegel}, D.~J. and {Secco}, L.~F. and {Vivas}, A.~K. and {Wester}, W. and {Zenteno}, A. and {Zhang}, Y. and {Abbott}, T.~M.~C. and {Banerji}, M. and {Bechtol}, K. and {Benoit-L{\'e}vy}, A. and {Bertin}, E. and {Buckley-Geer}, E. and {Burke}, D.~L. and {Capozzi}, D. and {Carnero Rosell}, A. and {Carrasco Kind}, M. and {Castander}, F.~J. and {Crocce}, M. and {Cunha}, C.~E. and {D'Andrea}, C.~B. and {da Costa}, L.~N. and {Davis}, C. and {DePoy}, D.~L. and {Desai}, S. and {Dietrich}, J.~P. and {Drlica-Wagner}, A. and {Eifler}, T.~F. and {Evrard}, A.~E. and {Fernandez}, E. and {Flaugher}, B. and {Fosalba}, P. and {Gaztanaga}, E. and {Gerdes}, D.~W. and {Giannantonio}, T. and {Goldstein}, D.~A. and {Gruen}, D. and {Gruendl}, R.~A. and {Gutierrez}, G. and {Honscheid}, K. and {Jain}, B. and {James}, D.~J. and {Jeltema}, T. and {Johnson}, M.~W.~G. and {Johnson}, M.~D. and {Kent}, S. and {Krause}, E. and {Kron}, R. and {Kuehn}, K. and {Nuropatkin}, N. and {Lahav}, O. and {Lima}, M. and {Lin}, H. and {Maia}, M.~A.~G. and {March}, M. and {Martini}, P. and {McMahon}, R.~G. and {Menanteau}, F. and {Miller}, C.~J. and {Miquel}, R. and {Mohr}, J.~J. and {Neilsen}, E. and {Nichol}, R.~C. and {Ogando}, R.~L.~C. and {Plazas}, A.~A. and {Roe}, N. and {Romer}, A.~K. and {Roodman}, A. and {Rykoff}, E.~S. and {Sanchez}, E. and {Scarpine}, V. and {Schindler}, R. and {Schubnell}, M. and {Sevilla-Noarbe}, I. and {Smith}, M. and {Smith}, R.~C. and {Sobreira}, F. and {Suchyta}, E. and {Swanson}, M.~E.~C. and {Tarle}, G. and {Thomas}, D. and {Thomas}, R.~C. and {Troxel}, M.~A. and {Vikram}, V. and {Walker}, A.~R. and {Wechsler}, R.~H. and {Weller}, J. and {Yanny}, B. and {Zuntz}, J.},
    title           = "{The Electromagnetic Counterpart of the Binary Neutron Star Merger LIGO/Virgo GW170817. II. UV, Optical, and Near-infrared Light Curves and Comparison to Kilonova Models}",
    journal         = {\apjl},
    year            = 2017,
    month           = oct,
    volume          = {848},
    number          = {2},
    eid             = {L17},
    pages           = {L17},
    doi             = {10.3847/2041-8213/aa8fc7},
    archivePrefix   = {arXiv},
    eprint          = {1710.05840},
    primaryClass    = {astro-ph.HE},
    adsurl          = {https://ui.adsabs.harvard.edu/abs/2017ApJ...848L..17C}
}

@ARTICLE{1996ApJ...469..794E,
       author = {{Engvik}, L. and {Osnes}, E. and {Hjorth-Jensen}, M. and {Bao}, G. and {Ostgaard}, E.},
        title = "{Asymmetric Nuclear Matter and Neutron Star Properties}",
      journal = {\apj},
         year = 1996,
        month = oct,
       volume = {469},
        pages = {794},
          doi = {10.1086/177827},
archivePrefix = {arXiv},
       eprint = {nucl-th/9509016},
 primaryClass = {nucl-th},
       adsurl = {https://ui.adsabs.harvard.edu/abs/1996ApJ...469..794E}
}

@ARTICLE{2016ARNPS..66...23F,
       author = {{Fern{\'a}ndez}, Rodrigo and {Metzger}, Brian D.},
        title = "{Electromagnetic Signatures of Neutron Star Mergers in the Advanced LIGO Era}",
      journal = {Annual Review of Nuclear and Particle Science},
         year = 2016,
        month = oct,
       volume = {66},
       number = {1},
        pages = {23-45},
          doi = {10.1146/annurev-nucl-102115-044819},
archivePrefix = {arXiv},
       eprint = {1512.05435},
 primaryClass = {astro-ph.HE},
       adsurl = {https://ui.adsabs.harvard.edu/abs/2016ARNPS..66...23F}
}

@ARTICLE{2020MNRAS.497.3221F,
       author = {{Fern{\'a}ndez}, Rodrigo and {Foucart}, Francois and {Lippuner}, Jonas},
        title = "{The landscape of disc outflows from black hole-neutron star mergers}",
      journal = {\mnras},
         year = 2020,
        month = sep,
       volume = {497},
       number = {3},
        pages = {3221-3233},
          doi = {10.1093/mnras/staa2209},
archivePrefix = {arXiv},
       eprint = {2005.14208},
 primaryClass = {astro-ph.HE},
       adsurl = {https://ui.adsabs.harvard.edu/abs/2020MNRAS.497.3221F}
}

@ARTICLE{2017CQGra..34d4002F,
       author = {{Foucart}, F. and {Desai}, D. and {Brege}, W. and {Duez}, M.~D. and {Kasen}, D. and {Hemberger}, D.~A. and {Kidder}, L.~E. and {Pfeiffer}, H.~P. and {Scheel}, M.~A.},
        title = "{Dynamical ejecta from precessing neutron star-black hole mergers with a hot, nuclear-theory based equation of state}",
      journal = {Classical and Quantum Gravity},
         year = 2017,
        month = feb,
       volume = {34},
       number = {4},
          eid = {044002},
        pages = {044002},
          doi = {10.1088/1361-6382/aa573b},
archivePrefix = {arXiv},
       eprint = {1611.01159},
 primaryClass = {astro-ph.HE},
       adsurl = {https://ui.adsabs.harvard.edu/abs/2017CQGra..34d4002F}
}

@ARTICLE{2018PhRvD..98h1501F,
       author = {{Foucart}, Francois and {Hinderer}, Tanja and {Nissanke}, Samaya},
        title = "{Remnant baryon mass in neutron star-black hole mergers: Predictions for binary neutron star mimickers and rapidly spinning black holes}",
      journal = {\prd},
         year = 2018,
        month = oct,
       volume = {98},
       number = {8},
          eid = {081501},
        pages = {081501},
          doi = {10.1103/PhysRevD.98.081501},
archivePrefix = {arXiv},
       eprint = {1807.00011},
 primaryClass = {astro-ph.HE},
       adsurl = {https://ui.adsabs.harvard.edu/abs/2018PhRvD..98h1501F}
}

@ARTICLE{2020PhRvD.101j3002K,
       author = {{Kr{\"u}ger}, Christian J. and {Foucart}, Francois},
        title = "{Estimates for disk and ejecta masses produced in compact binary mergers}",
      journal = {\prd},
         year = 2020,
        month = may,
       volume = {101},
       number = {10},
          eid = {103002},
        pages = {103002},
          doi = {10.1103/PhysRevD.101.103002},
archivePrefix = {arXiv},
       eprint = {2002.07728},
 primaryClass = {astro-ph.HE},
       adsurl = {https://ui.adsabs.harvard.edu/abs/2020PhRvD.101j3002K}
}

@ARTICLE{2001ApJ...550..426L,
       author = {{Lattimer}, J.~M. and {Prakash}, M.},
        title = "{Neutron Star Structure and the Equation of State}",
      journal = {\apj},
         year = 2001,
        month = mar,
       volume = {550},
       number = {1},
        pages = {426-442},
          doi = {10.1086/319702},
archivePrefix = {arXiv},
       eprint = {astro-ph/0002232},
 primaryClass = {astro-ph},
       adsurl = {https://ui.adsabs.harvard.edu/abs/2001ApJ...550..426L}
}

@ARTICLE{2017ApJS..233...18L,
       author = {{Lippuner}, Jonas and {Roberts}, Luke F.},
        title = "{SkyNet: A Modular Nuclear Reaction Network Library}",
      journal = {\apjs},
         year = 2017,
        month = dec,
       volume = {233},
       number = {2},
          eid = {18},
        pages = {18},
          doi = {10.3847/1538-4365/aa94cb},
archivePrefix = {arXiv},
       eprint = {1706.06198},
 primaryClass = {astro-ph.HE},
       adsurl = {https://ui.adsabs.harvard.edu/abs/2017ApJS..233...18L}
}

@ARTICLE{2015MNRAS.452.1970W,
       author = {{Wehmeyer}, B. and {Pignatari}, M. and {Thielemann}, F.-K.},
        title = "{Galactic evolution of rapid neutron capture process abundances: the inhomogeneous approach}",
      journal = {\mnras},
         year = 2015,
        month = sep,
       volume = {452},
       number = {2},
        pages = {1970-1981},
          doi = {10.1093/mnras/stv1352},
archivePrefix = {arXiv},
       eprint = {1501.07749},
 primaryClass = {astro-ph.GA},
       adsurl = {https://ui.adsabs.harvard.edu/abs/2015MNRAS.452.1970W}
}

@ARTICLE{2014ARA&A..52..415M,
       author = {{Madau}, Piero and {Dickinson}, Mark},
        title = "{Cosmic Star-Formation History}",
      journal = {\araa},
         year = 2014,
        month = aug,
       volume = {52},
        pages = {415-486},
          doi = {10.1146/annurev-astro-081811-125615},
archivePrefix = {arXiv},
       eprint = {1403.0007},
 primaryClass = {astro-ph.CO},
       adsurl = {https://ui.adsabs.harvard.edu/abs/2014ARA&A..52..415M}
}

@ARTICLE{2021MNRAS.505..339M,
       author = {{Mapelli}, Michela and {Dall'Amico}, Marco and {Bouffanais}, Yann and {Giacobbo}, Nicola and {Arca Sedda}, Manuel and {Artale}, M. Celeste and {Ballone}, Alessandro and {Di Carlo}, Ugo N. and {Iorio}, Giuliano and {Santoliquido}, Filippo and {Torniamenti}, Stefano},
        title = "{Hierarchical black hole mergers in young, globular and nuclear star clusters: the effect of metallicity, spin and cluster properties}",
      journal = {\mnras},
         year = 2021,
        month = jul,
       volume = {505},
       number = {1},
        pages = {339-358},
          doi = {10.1093/mnras/stab1334},
archivePrefix = {arXiv},
       eprint = {2103.05016},
 primaryClass = {astro-ph.HE},
       adsurl = {https://ui.adsabs.harvard.edu/abs/2021MNRAS.505..339M}
}

@ARTICLE{2006PASJ...58..847N,
       author = {{Nakanishi}, Hiroyuki and {Sofue}, Yoshiaki},
        title = "{Three-Dimensional Distribution of the ISM in the Milky Way Galaxy: II. The Molecular Gas Disk}",
      journal = {\pasj},
         year = 2006,
        month = oct,
       volume = {58},
        pages = {847-860},
          doi = {10.1093/pasj/58.5.847},
archivePrefix = {arXiv},
       eprint = {astro-ph/0610769},
 primaryClass = {astro-ph},
       adsurl = {https://ui.adsabs.harvard.edu/abs/2006PASJ...58..847N}
}

@ARTICLE{2022CQGra..39a5008N,
       author = {{Nedora}, Vsevolod and {Schianchi}, Federico and {Bernuzzi}, Sebastiano and {Radice}, David and {Daszuta}, Boris and {Endrizzi}, Andrea and {Perego}, Albino and {Prakash}, Aviral and {Zappa}, Francesco},
        title = "{Mapping dynamical ejecta and disk masses from numerical relativity simulations of neutron star mergers}",
      journal = {Classical and Quantum Gravity},
         year = 2022,
        month = jan,
       volume = {39},
       number = {1},
          eid = {015008},
        pages = {015008},
          doi = {10.1088/1361-6382/ac35a8},
archivePrefix = {arXiv},
       eprint = {2011.11110},
 primaryClass = {astro-ph.HE},
       adsurl = {https://ui.adsabs.harvard.edu/abs/2022CQGra..39a5008N}
}

@ARTICLE{2017ApJ...848L..18N,
       author = {{Nicholl}, M. and {Berger}, E. and {Kasen}, D. and {Metzger}, B.~D. and {Elias}, J. and {Brice{\~n}o}, C. and {Alexander}, K.~D. and {Blanchard}, P.~K. and {Chornock}, R. and {Cowperthwaite}, P.~S. and {Eftekhari}, T. and {Fong}, W. and {Margutti}, R. and {Villar}, V.~A. and {Williams}, P.~K.~G. and {Brown}, W. and {Annis}, J. and {Bahramian}, A. and {Brout}, D. and {Brown}, D.~A. and {Chen}, H.-Y. and {Clemens}, J.~C. and {Dennihy}, E. and {Dunlap}, B. and {Holz}, D.~E. and {Marchesini}, E. and {Massaro}, F. and {Moskowitz}, N. and {Pelisoli}, I. and {Rest}, A. and {Ricci}, F. and {Sako}, M. and {Soares-Santos}, M. and {Strader}, J.},
        title = "{The Electromagnetic Counterpart of the Binary Neutron Star Merger LIGO/Virgo GW170817. III. Optical and UV Spectra of a Blue Kilonova from Fast Polar Ejecta}",
      journal = {\apjl},
         year = 2017,
        month = oct,
       volume = {848},
       number = {2},
          eid = {L18},
        pages = {L18},
          doi = {10.3847/2041-8213/aa9029},
archivePrefix = {arXiv},
       eprint = {1710.05456},
 primaryClass = {astro-ph.HE},
       adsurl = {https://ui.adsabs.harvard.edu/abs/2017ApJ...848L..18N}
}

@ARTICLE{1964PhRv..136.1224P,
       author = {{Peters}, P.~C.},
        title = "{Gravitational Radiation and the Motion of Two Point Masses}",
      journal = {Physical Review},
         year = 1964,
        month = nov,
       volume = {136},
       number = {4B},
        pages = {1224-1232},
          doi = {10.1103/PhysRev.136.B1224},
       adsurl = {https://ui.adsabs.harvard.edu/abs/1964PhRv..136.1224P}
}

@ARTICLE{2018ApJ...869..130R,
       author = {{Radice}, David and {Perego}, Albino and {Hotokezaka}, Kenta and {Fromm}, Steven A. and {Bernuzzi}, Sebastiano and {Roberts}, Luke F.},
        title = "{Binary Neutron Star Mergers: Mass Ejection, Electromagnetic Counterparts, and Nucleosynthesis}",
      journal = {\apj},
         year = 2018,
        month = dec,
       volume = {869},
       number = {2},
          eid = {130},
        pages = {130},
          doi = {10.3847/1538-4357/aaf054},
archivePrefix = {arXiv},
       eprint = {1809.11161},
 primaryClass = {astro-ph.HE},
       adsurl = {https://ui.adsabs.harvard.edu/abs/2018ApJ...869..130R}
}

@ARTICLE{2014AJ....147..136R,
       author = {{Roederer}, Ian U. and {Preston}, George W. and {Thompson}, Ian B. and {Shectman}, Stephen A. and {Sneden}, Christopher and {Burley}, Gregory S. and {Kelson}, Daniel D.},
        title = "{A Search for Stars of Very Low Metal Abundance. VI. Detailed Abundances of 313 Metal-poor Stars}",
      journal = {\aj},
         year = 2014,
        month = jun,
       volume = {147},
       number = {6},
          eid = {136},
        pages = {136},
          doi = {10.1088/0004-6256/147/6/136},
archivePrefix = {arXiv},
       eprint = {1403.6853},
 primaryClass = {astro-ph.SR},
       adsurl = {https://ui.adsabs.harvard.edu/abs/2014AJ....147..136R}
}

@ARTICLE{2013ApJ...774...17S,
       author = {{Steiner}, A.~W. and {Hempel}, M. and {Fischer}, T.},
        title = "{Core-collapse Supernova Equations of State Based on Neutron Star Observations}",
      journal = {\apj},
         year = 2013,
        month = sep,
       volume = {774},
       number = {1},
          eid = {17},
        pages = {17},
          doi = {10.1088/0004-637X/774/1/17},
archivePrefix = {arXiv},
       eprint = {1207.2184},
 primaryClass = {astro-ph.SR},
       adsurl = {https://ui.adsabs.harvard.edu/abs/2013ApJ...774...17S}
}

@ARTICLE{2020MNRAS.494.4867V,
       author = {{van de Voort}, Freeke and {Pakmor}, R{\"u}diger and {Grand}, Robert J.~J. and {Springel}, Volker and {G{\'o}mez}, Facundo A. and {Marinacci}, Federico},
        title = "{Neutron star mergers and rare core-collapse supernovae as sources of r-process enrichment in simulated galaxies}",
      journal = {\mnras},
         year = 2020,
        month = jun,
       volume = {494},
       number = {4},
        pages = {4867-4883},
          doi = {10.1093/mnras/staa754},
archivePrefix = {arXiv},
       eprint = {1907.01557},
 primaryClass = {astro-ph.GA},
       adsurl = {https://ui.adsabs.harvard.edu/abs/2020MNRAS.494.4867V}
}

@ARTICLE{2010PhRvC..81a5803T,
       author = {{Typel}, S. and {R{\"o}pke}, G. and {Kl{\"a}hn}, T. and {Blaschke}, D. and {Wolter}, H.~H.},
        title = "{Composition and thermodynamics of nuclear matter with light clusters}",
      journal = {\prc},
         year = 2010,
        month = jan,
       volume = {81},
       number = {1},
          eid = {015803},
        pages = {015803},
          doi = {10.1103/PhysRevC.81.015803},
archivePrefix = {arXiv},
       eprint = {0908.2344},
 primaryClass = {nucl-th},
       adsurl = {https://ui.adsabs.harvard.edu/abs/2010PhRvC..81a5803T}
}

@ARTICLE{1984ApJ...277..355W,
       author = {{Webbink}, R.~F.},
        title = "{Double white dwarfs as progenitors of R Coronae Borealis stars and type I supernovae.}",
      journal = {\apj},
         year = 1984,
        month = feb,
       volume = {277},
        pages = {355-360},
          doi = {10.1086/161701},
       adsurl = {https://ui.adsabs.harvard.edu/abs/1984ApJ...277..355W}
}

@ARTICLE{2021ApJ...919...59C,
       author = {{Chen}, Meng-Hua and {Li}, Li-Xin and {Lin}, Da-Bin and {Liang}, En-Wei},
        title = "{Gamma-Ray Emission Produced by r-process Elements from Neutron Star Mergers}",
      journal = {\apj},
         year = 2021,
        month = sep,
       volume = {919},
       number = {1},
          eid = {59},
        pages = {59},
          doi = {10.3847/1538-4357/ac1267},
archivePrefix = {arXiv},
       eprint = {2107.02982},
 primaryClass = {astro-ph.HE},
       adsurl = {https://ui.adsabs.harvard.edu/abs/2021ApJ...919...59C}
}

@ARTICLE{2024MNRAS.529.1154C,
       author = {{Chen}, Meng-Hua and {Li}, Li-Xin and {Chen}, Qiu-Hong and {Hu}, Rui-Chong and {Liang}, En-Wei},
        title = "{Neutron star mergers as the dominant contributor to the production of heavy r-process elements}",
      journal = {\mnras},
         year = 2024,
        month = apr,
       volume = {529},
       number = {2},
        pages = {1154-1160},
          doi = {10.1093/mnras/stae475},
archivePrefix = {arXiv},
       eprint = {2402.08214},
 primaryClass = {astro-ph.HE},
       adsurl = {https://ui.adsabs.harvard.edu/abs/2024MNRAS.529.1154C}
}

@ARTICLE{2015PhRvD..91l4021F,
       author = {{Foucart}, Francois and {O'Connor}, Evan and {Roberts}, Luke and {Duez}, Matthew D. and {Haas}, Roland and {Kidder}, Lawrence E. and {Ott}, Christian D. and {Pfeiffer}, Harald P. and {Scheel}, Mark A. and {Szilagyi}, Bela},
        title = "{Post-merger evolution of a neutron star-black hole binary with neutrino transport}",
      journal = {\prd},
         year = 2015,
        month = jun,
       volume = {91},
       number = {12},
          eid = {124021},
        pages = {124021},
          doi = {10.1103/PhysRevD.91.124021},
archivePrefix = {arXiv},
       eprint = {1502.04146},
 primaryClass = {astro-ph.HE},
       adsurl = {https://ui.adsabs.harvard.edu/abs/2015PhRvD..91l4021F}
}

@ARTICLE{2015MNRAS.449..390F,
       author = {{Fern{\'a}ndez}, Rodrigo and {Quataert}, Eliot and {Schwab}, Josiah and {Kasen}, Daniel and {Rosswog}, Stephan},
        title = "{The interplay of disc wind and dynamical ejecta in the aftermath of neutron star-black hole mergers}",
      journal = {\mnras},
         year = 2015,
        month = may,
       volume = {449},
       number = {1},
        pages = {390-402},
          doi = {10.1093/mnras/stv238},
archivePrefix = {arXiv},
       eprint = {1412.5588},
 primaryClass = {astro-ph.HE},
       adsurl = {https://ui.adsabs.harvard.edu/abs/2015MNRAS.449..390F}
}

@ARTICLE{2023PhRvD.107l3001H,
       author = {{Hayashi}, Kota and {Kiuchi}, Kenta and {Kyutoku}, Koutarou and {Sekiguchi}, Yuichiro and {Shibata}, Masaru},
        title = "{General-relativistic neutrino-radiation magnetohydrodynamics simulation of seconds-long black hole-neutron star mergers: Dependence on the initial magnetic field strength, configuration, and neutron-star equation of state}",
      journal = {\prd},
         year = 2023,
        month = jun,
       volume = {107},
       number = {12},
          eid = {123001},
        pages = {123001},
          doi = {10.1103/PhysRevD.107.123001},
archivePrefix = {arXiv},
       eprint = {2211.07158},
 primaryClass = {astro-ph.HE},
       adsurl = {https://ui.adsabs.harvard.edu/abs/2023PhRvD.107l3001H}
}

@ARTICLE{2001AJ....121..753B,
       author = {{Boselli}, A. and {Gavazzi}, G. and {Donas}, J. and {Scodeggio}, M.},
        title = "{1.65 Micron (H Band) Surface Photometry of Galaxies. VI. The History of Star Formation in Normal Late-Type Galaxies}",
      journal = {\aj},
         year = 2001,
        month = feb,
       volume = {121},
       number = {2},
        pages = {753-767},
          doi = {10.1086/318734},
archivePrefix = {arXiv},
       eprint = {astro-ph/0011016},
 primaryClass = {astro-ph},
       adsurl = {https://ui.adsabs.harvard.edu/abs/2001AJ....121..753B}
}

@ARTICLE{1986A&A...161...89S,
       author = {{Sandage}, A.},
        title = "{Star formation rates, galaxy morphology and the Hubble sequence.}",
      journal = {\aap},
         year = 1986,
        month = jun,
       volume = {161},
        pages = {89-101},
       adsurl = {https://ui.adsabs.harvard.edu/abs/1986A&A...161...89S}
}

@ARTICLE{2008MNRAS.384..267M,
       author = {{Maoz}, Dan},
        title = "{On the fraction of intermediate-mass close binaries that explode as Type Ia supernovae}",
      journal = {\mnras},
         year = 2008,
        month = feb,
       volume = {384},
       number = {1},
        pages = {267-277},
          doi = {10.1111/j.1365-2966.2007.12697.x},
archivePrefix = {arXiv},
       eprint = {0707.4598},
 primaryClass = {astro-ph},
       adsurl = {https://ui.adsabs.harvard.edu/abs/2008MNRAS.384..267M}
}

@ARTICLE{2024ApJ...973..122W,
       author = {{Weinberg}, David H. and {Griffith}, Emily J. and {Johnson}, James W. and {Thompson}, Todd A.},
        title = "{The Scale of Stellar Yields: Implications of the Measured Mean Iron Yield of Core Collapse Supernovae}",
      journal = {\apj},
         year = 2024,
        month = oct,
       volume = {973},
       number = {2},
          eid = {122},
        pages = {122},
          doi = {10.3847/1538-4357/ad6313},
archivePrefix = {arXiv},
       eprint = {2309.05719},
 primaryClass = {astro-ph.GA},
       adsurl = {https://ui.adsabs.harvard.edu/abs/2024ApJ...973..122W}
}

@ARTICLE{1998ApJ...500...95T,
       author = {{Thornton}, K. and {Gaudlitz}, M. and {Janka}, H.-Th. and {Steinmetz}, M.},
        title = "{Energy Input and Mass Redistribution by Supernovae in the Interstellar Medium}",
      journal = {\apj},
         year = 1998,
        month = jun,
       volume = {500},
       number = {1},
        pages = {95-119},
          doi = {10.1086/305704},
archivePrefix = {arXiv},
       eprint = {astro-ph/9706175},
 primaryClass = {astro-ph},
       adsurl = {https://ui.adsabs.harvard.edu/abs/1998ApJ...500...95T}
}

@ARTICLE{2014ARA&A..52..107M,
       author = {{Maoz}, Dan and {Mannucci}, Filippo and {Nelemans}, Gijs},
        title = "{Observational Clues to the Progenitors of Type Ia Supernovae}",
      journal = {\araa},
         year = 2014,
        month = aug,
       volume = {52},
        pages = {107-170},
          doi = {10.1146/annurev-astro-082812-141031},
archivePrefix = {arXiv},
       eprint = {1312.0628},
 primaryClass = {astro-ph.CO},
       adsurl = {https://ui.adsabs.harvard.edu/abs/2014ARA&A..52..107M}
}

@ARTICLE{2001MNRAS.322..231K,
       author = {{Kroupa}, Pavel},
        title = "{On the variation of the initial mass function}",
      journal = {\mnras},
         year = 2001,
        month = apr,
       volume = {322},
       number = {2},
        pages = {231-246},
          doi = {10.1046/j.1365-8711.2001.04022.x},
archivePrefix = {arXiv},
       eprint = {astro-ph/0009005},
 primaryClass = {astro-ph},
       adsurl = {https://ui.adsabs.harvard.edu/abs/2001MNRAS.322..231K}
}

@ARTICLE{2009ApJ...698.1963R,
       author = {{Roederer}, Ian U. and {Kratz}, Karl-Ludwig and {Frebel}, Anna and {Christlieb}, Norbert and {Pfeiffer}, Bernd and {Cowan}, John J. and {Sneden}, Christopher},
        title = "{The End of Nucleosynthesis: Production of Lead and Thorium in the Early Galaxy}",
      journal = {\apj},
         year = 2009,
        month = jun,
       volume = {698},
       number = {2},
        pages = {1963-1980},
          doi = {10.1088/0004-637X/698/2/1963},
archivePrefix = {arXiv},
       eprint = {0904.3105},
 primaryClass = {astro-ph.SR},
       adsurl = {https://ui.adsabs.harvard.edu/abs/2009ApJ...698.1963R}
}

@ARTICLE{2004AJ....128.1177V,
       author = {{Venn}, Kim A. and {Irwin}, Mike and {Shetrone}, Matthew D. and {Tout}, Christopher A. and {Hill}, Vanessa and {Tolstoy}, Eline},
        title = "{Stellar Chemical Signatures and Hierarchical Galaxy Formation}",
      journal = {\aj},
         year = 2004,
        month = sep,
       volume = {128},
       number = {3},
        pages = {1177-1195},
          doi = {10.1086/422734},
archivePrefix = {arXiv},
       eprint = {astro-ph/0406120},
 primaryClass = {astro-ph},
       adsurl = {https://ui.adsabs.harvard.edu/abs/2004AJ....128.1177V}
}

@ARTICLE{2016A&A...586A..49B,
       author = {{Battistini}, Chiara and {Bensby}, Thomas},
        title = "{The origin and evolution of r- and s-process elements in the Milky Way stellar disk}",
      journal = {\aap},
         year = 2016,
        month = feb,
       volume = {586},
          eid = {A49},
        pages = {A49},
          doi = {10.1051/0004-6361/201527385},
archivePrefix = {arXiv},
       eprint = {1511.00966},
 primaryClass = {astro-ph.SR},
       adsurl = {https://ui.adsabs.harvard.edu/abs/2016A&A...586A..49B}
}

@ARTICLE{2015ApJ...807..171J,
       author = {{Jacobson}, Heather R. and {Keller}, Stefan and {Frebel}, Anna and {Casey}, Andrew R. and {Asplund}, Martin and {Bessell}, Michael S. and {Da Costa}, Gary S. and {Lind}, Karin and {Marino}, Anna F. and {Norris}, John E. and {Pe{\~n}a}, Jos{\'e} M. and {Schmidt}, Brian P. and {Tisserand}, Patrick and {Walsh}, Jennifer M. and {Yong}, David and {Yu}, Qinsi},
        title = "{High-Resolution Spectroscopic Study of Extremely Metal-Poor Star Candidates from the SkyMapper Survey}",
      journal = {\apj},
         year = 2015,
        month = jul,
       volume = {807},
       number = {2},
          eid = {171},
        pages = {171},
          doi = {10.1088/0004-637X/807/2/171},
archivePrefix = {arXiv},
       eprint = {1504.03344},
 primaryClass = {astro-ph.SR},
       adsurl = {https://ui.adsabs.harvard.edu/abs/2015ApJ...807..171J}
}

@ARTICLE{2017ApJ...835..128C,
       author = {{C{\^o}t{\'e}}, Benoit and {O'Shea}, Brian W. and {Ritter}, Christian and {Herwig}, Falk and {Venn}, Kim A.},
        title = "{The Impact of Modeling Assumptions in Galactic Chemical Evolution Models}",
      journal = {\apj},
         year = 2017,
        month = feb,
       volume = {835},
       number = {2},
          eid = {128},
        pages = {128},
          doi = {10.3847/1538-4357/835/2/128},
archivePrefix = {arXiv},
       eprint = {1604.07824},
 primaryClass = {astro-ph.GA},
       adsurl = {https://ui.adsabs.harvard.edu/abs/2017ApJ...835..128C}
}

@ARTICLE{2018ApJ...860...89M,
       author = {{Macias}, Phillip and {Ramirez-Ruiz}, Enrico},
        title = "{A Stringent Limit on the Mass Production Rate of r-process Elements in the Milky Way}",
      journal = {\apj},
         year = 2018,
        month = jun,
       volume = {860},
       number = {2},
          eid = {89},
        pages = {89},
          doi = {10.3847/1538-4357/aac3e0},
archivePrefix = {arXiv},
       eprint = {1609.04826},
 primaryClass = {astro-ph.HE},
       adsurl = {https://ui.adsabs.harvard.edu/abs/2018ApJ...860...89M}
}

@BOOK{2018pgrb.book.....Z,
       author = {{Zhang}, Bing},
        title = "{The Physics of Gamma-Ray Bursts}",
         year = 2018,
          doi = {10.1017/9781139226530},
       adsurl = {https://ui.adsabs.harvard.edu/abs/2018pgrb.book.....Z}
}
\end{document}